\documentclass{article}

\usepackage{PRIMEarxiv}
\usepackage{fancyhdr}       
\usepackage{tikz}
\newlength\imageheight
\usepackage{amsfonts}
\usepackage{subfig}
\usepackage{amsmath}
\usepackage{bm}
\usepackage{enumitem}
\usepackage{amssymb}
\usepackage{makecell}
\usepackage{caption} 
\usepackage{array,multirow,graphicx}
\usepackage{float}
\usepackage[ruled,vlined, linesnumbered]{algorithm2e}
\usepackage{hyperref}
\usepackage{authblk}

\title{Modeling Teams Performance Using Deep Representational Learning on Graphs \thanks{This work has been submitted to the IEEE for possible publication. Copyright may be transferred without notice, after which this version may no longer be accessible.}}

\author[1,2]{Francesco Carli}
\author[1]{Pietro Foini}
\author[3]{Nicolò Gozzi}
\author[3,4]{Nicola Perra}
\author[1,5]{Rossano Schifanella}
\affil[1]{Computer Science Department, University of Turin}
\affil[2]{Bioinformatics Lab, Scuola Normale Superiore, Pisa}
\affil[3]{Networks and Urban Systems Centre, University of Greenwich}
\affil[4]{School of Mathematical Sciences, Queen University of London}
\affil[5]{ISI Foundation, Turin}

\begin{document}
\maketitle

\begin{abstract}
The large majority of human activities require collaborations within and across formal or informal teams. Our understanding of how the collaborative efforts spent by teams relate to their performance is still a matter of debate. Teamwork results into a highly interconnected ecosystem of potentially overlapping components where tasks are performed in interaction with team members and across other teams.
To tackle this problem, we propose a graph neural network model designed to predict a team's performance while identifying the drivers that determine such outcome. In particular, the model is based on three architectural channels: topological, centrality and contextual which capture different factors potentially shaping teams' success. We endow the model with two attention mechanisms to boost model performance and allow interpretability. A first mechanism allows pinpointing key members inside the team. A second mechanism allows us to quantify the contributions of the three driver effects in determining the outcome performance. We test model performance on a wide range of domains outperforming most of the classical and neural baselines considered. Moreover, we include synthetic datasets specifically designed to validate how the model disentangles the intended properties on which our model vastly outperforms baselines. 
\end{abstract}


\section{Introduction}
\label{sec:introduction}

What makes a team effective is a long-standing problem widely studied across disciplines and applicative contexts. Several factors such as communication, coordination, distinctive roles, interdependent tasks, shared norms, personality traits, and diversity have been found to be relevant aspects shaping team performance~\cite{ducanis1979interdisciplinary,Brannick1997TeamPA,Peeters,Bell2011,pentland2012new,NYT}. Yet, our understanding of teams as evolving systems of interacting individuals as well as the relation between team composition and performance is still partial~\cite{delice2019advancing,Mathieu,carter2015little}. 
 
When studying teams a key issue is how to combine the features (e.g., skills, socio-demographic indicators, relations, and past experiences) of single individuals at the team level. Straightforward solutions are offered by the so-called \textit{compositional} models~\cite{Kozlowski}. They rely on the assumption that the contribution of each team member is equal. As a result, attributes of single individuals are considered additive and possibly averaged in a summary index~\cite{Kozlowski}. However, this approach provides an extreme simplification of the dynamics at play. In contrast, in \textit{compilational} models, team-level attributes are considered complex combinations of individual-level properties~\cite{Kozlowski}. The intuition is that teams could be more than the sum of their parts. Perspectives from Complexity and Network Science offer natural frameworks to capture and investigate this direction~\cite{merton1968matthew, barabasi1999emergence,yucesoy2016untangling,mcpherson2001birds,kossinets2009origins,ducanis1979interdisciplinary,Brannick1997TeamPA,Peeters,Bell2011,pentland2012new, bell2018}. Within these approaches, teams' performance has been linked to three effects. The first are \textit{topological} effects. The internal structure of a team, emerging from the interactions of its members, plays a crucial role in determining performance~\cite{ducanis1979interdisciplinary,Brannick1997TeamPA,Peeters,Bell2011,pentland2012new,guimera}. The second are \textit{centrality} effects. Teams' performance is influenced by importance/role of a team with respect to the ecosystem which it belongs to. Indeed, collaborations (i.e., connections) with people outside the team, sharing and advertising of one's work are key factors that might boost teams' performances by leveraging popularity, rich get richer phenomena, and provide access to relevant as well as novel information~\cite{merton1968matthew, barabasi1999emergence,yucesoy2016untangling}. Centrality effects, commonly encoded through network metrics such as degree, betweenness and closeness~\cite{sabidussi1966centrality}, roughly capture the overall visibility of a team in the system as well as its ability to be part of informative flows. The third are \textit{contextual} effects. The success or failure of a team can be guided by the context to which it belongs and in which it develops, regardless of how internal or external relations are structured. For example, the number of citations received by articles published by research teams might vary significantly across different disciplines~\cite{impact}. It depends on the context where the activity, such as a publication, takes place.

The identification of the drivers of teams performance is an important step but does not solve the problem. Indeed, the hand-design of features that allow models to capture the complex effects of such factors is far from trivial. Recent advancements in the extension of deep learning architectures to graph structured data can help us solve this challenge~\cite{GCN,GraphSAGE,GAT,GIN}. \textit{Graph Neural Networks} (GNNs) offers a natural way to derive high-order representations of interacting systems by inferring, in this application, the relevant and holistic features of the team as a result of a learning procedure. In this regard, the interacting systems of interest here fit well in a graph-based scenario. The whole graph represents the ecosystem in which the collaborative activity takes place and the teams are represented by their parts (i.e., subgraphs). Therefore, the task of modeling team performance can be rephrased in terms of designing graph representation learning methods able to project the subgraph structures into a higher dimensional space, called embedding space, that can be subsequently leveraged by a downstream classifier to solve tasks of interest.

Learning methods on graphs greatly improved in recent years~\cite{SurveyGNN}. However, the literature on GNNs aims at developing architectures useful in learning representations for nodes~\cite{GCN,GraphSAGE,GAT}, edges~\cite{gao2019edge2vec, edges2} or entire graphs~\cite{GIN}. Therefore, these methodologies may not be optimal in modeling the broad spectrum of teams (i.e., subgraphs) peculiarities. As highlighted in Alsentenzer et al.~\cite{SUBGNN}, subgraphs have non-trivial internal structure, border connectivity, and notions of neighborhood and position relative to the rest of the graph. Therefore, tackling the problem of subgraph embedding requires the design of architectures able to capture graph features that may not be defined for finer and coarser graph components such as nodes or whole graphs. 

Here, we present a compilational model, based on graph neural learning, that aims at capturing the dynamics that shape team performance. The model explicitly considers for \textit{topological}, \textit{centrality} and \textit{contextual} effects. We summarize our contributions as follows: 

\begin{itemize}
    \item We propose \textit{MENTOR} (\textbf{M}od\textbf{E}li\textbf{N}g \textbf{T}eams Performance Using Deep Representational Learning \textbf{O}n G\textbf{R}aphs), a new three channels architecture (Figure~\ref{fig:Framework}) that models team performance by leveraging topological, centrality and contextual effects. In more depth, this architecture features graph neural learning methods defined on subgraph structures;
    \item We endow the model with two attention mechanisms that allow us to examine in more detail targeted parts of the proposed deep architecture. A first mechanism, defined at the node level, allows pinpointing key members inside the team. A second mechanism, defined at the channels aggregation level, allows us to quantify the contributions of topological, centrality and contextual effects in determining the final outcome. These two mechanisms not only enhance the model's expressivity but also shed light on the inner workings of the proposed architecture, providing some degree of interpretability;  
    \item We test the model's performance on a wide range of domains. Furthermore, we introduce synthetic datasets specifically designed to include topological, centrality and contextual effects. This allows us to test whether the proposed architecture is capable to learn disentangled representations of the intended properties. We than show how the proposed model outperforms most of of classical and neural baselines on the analyzed datasets.
\end{itemize}

\section{Related work}

An extensive body of research has focused on what are the key factors that affect team performance. Works from a range of disciplines identified features like regular communication, coordination, distinctive roles, interdependent tasks and shared norms as the building blocks of effective teams~\cite{ducanis1979interdisciplinary, Brannick1997TeamPA, Peeters, Bell2011, Humphrey}. 

Several models have been proposed and evaluated in different contexts. For example, the seminal work by McGrath~\cite{McGrath} introduced an input-process-output (IPO) model where antecedent conditions and resources (i.e., input) maintain internal processes and produce specific products (i.e., output). According to this model, the necessary antecedent conditions together with the processes of maintaining teams define their effectiveness. A relevant body of literature is attributable to this paradigm and its extensions; however, it has been shown to be too simplistic and unable to accurately account for all the complex interactions that influence team performance~\cite{forsyth2009group}. 

More in general, research on team composition focuses on the attributes of team members and the impact of their combination on processes, emergent states, and ultimately performance~\cite{Mathieu}. The research on the subject can be grouped into three main areas~\cite{levine2006small}: i) studies focusing on the features of team members, ii) studies focusing on how such features are measured, and iii) studies that investigate alternative approaches to team composition. As part of the last category, Kozlowski and Klein~\cite{Kozlowski} described composition processes as relatively simple combination rules aimed to shift from lower-level units (i.e., individual) to higher-level constructs (i.e., team-level attributes). Two main general approaches are commonly used to describe team composition. The first category considers compositional models. As mentioned above, these assume that members are “isomorphic” and that their contribution is equally weighted. Examples are models based on mean and diversity indices. The former computes team-level scores as the mean of the individual-level attributes~\cite{Chen2004}. This is the so-called “all-stars” approach. In fact, it implies that the best teams are those formed by ensembles of top individuals. The latter, instead, assumes that the heterogeneity (i.e., diversity) of the attributes at the lower level is crucial and it is often operationalized with measures like variance~\cite{Chen2004}. In general, compositional models are simplistic. They do not capture the dynamics at play and do not apply to tasks/contexts where individual contributions are not as significant as teamwork. An example comes from sports, where an “all-star” team is not necessarily the best. Furthermore, empirical findings show that team performance is not a monotonic function of diversity~\cite{uzzi}. The second category considers compilational models. The overarching assumption is that team-level attributes cannot be computed from simple statistical measures of lower-level quantities. Teamwork implies interactions between members and thus between their attributes. Within this vision, teams are considered as complex adaptive systems made up of multiple parts that continually interact and adapt their behavior in response to the behavior of the other parts~\cite{arrow}. Thus, compilational models consider complex combinations of members attributes such as the relative position or status of the highest or lowest individual~\cite{Bell2007, Mathieu} or network features that capture the structural properties of the social connections linking members between and within teams~\cite{borgatti, guimera}. By modeling such systems, researchers seek to understand how the aggregate behavior emerges from the interactions of the parts, integrating multiple levels of analysis to build a more thorough understanding~\cite{ramos}. 

From a methodological point of view, Machine Learning (ML) has been recently applied to the context of team modeling in a wide range of applicative scenarios, mainly with cross-sectional data and hand-designed features carefully engineered by domain experts. In particular, we refer to the example of online games~\cite{sapienza, goyal, Chen2019} where interactions and performances of teams are captured through real-time data collection platforms. Similar to the goals of this work, Chen at al.~\cite{Chen2019} aimed at understanding what makes a good team in Honor Kings, a massive online game with more than 96 million users. However, they limit their analysis by i) focusing on specific aspects while overlooking the holistic picture underpinning team dynamics ii) adopting hand-designed features that often fail to capture the complexity of high-order functional relationships. All these limitations clearly pointed out how these approaches should be refined to unravel the complex threads behind these scenarios.

Recent literature on graph representation learning shows how deep learning architectures and hence implicit feature engineering can be achieved by designing methods that, without hand-crafted features, capture patterns of compound interactions. In particular, these methods deal with graph-structured input data. In more depth, several network embedding frameworks were proposed to represent graph nodes as low-dimensional vectors~\cite{hamilton2020graph,perozzi2014deepwalk,grover2016node2vec,liao2018attributed}. Such representations aim at preserving both network topology structure and node features, delivering embeddings that can be used downstream for tasks such as classification, clustering, and link prediction through classical machine learning methods.
In this work we will focus on a class of models broadly referred to as \textit{graph neural networks} (GNNs) \cite{scarselli2008graph,micheli2009neural}. GNNs perform neighborhood aggregation through a procedure called \textit{neural message-passing} \cite{gilmer2017neural}, where the embeddings of nodes are obtained by recursively pooling and transforming representation vectors of their neighborhood~\cite{GCN,GraphSAGE,GAT,GIN}. Despite works such as references~\cite{battaglia2018relational,chami2020machine}, which try to outline a general unifying GNN framework for all of the applications introduced so far, deep learning on graphs is a fast-evolving field, and many theoretical results still need to be proved. A fair amount of recent research~\cite{GIN,xu2019can,maron2019provably} involves shedding light on GNNs \textit{expressivity}, with particular focus on understanding: a) the relation between the \textit{depth} of a GNN architecture and \textit{over smoothing} \cite{jumping,li2018deeper,klicpera2018predict}, b) the interplay of \textit{positional} and \textit{structural effects} \cite{you2019position,srinivasan2019equivalence}, c) the difference between \textit{homophily} and \textit{heterophily} in graphs \cite{yan2021two,bo2021beyond}.

Most of the introduced literature on GNNs focuses mainly on node-level or graph-level tasks~\cite{GraphSAGE,GAT,GIN}. As we will explain in depth later, team modeling involves a subgraph learning problem. Subgraph embedding tasks by means of GNNs are still an underexplored area of research with SubGNN model~\cite{SUBGNN} being the most notable exception. Similar to the framework that we propose, SubGNN tackles the problem of embedding general subgraphs by specifying three channels designed to capture distinct aspects of subgraph structures. However, being SubGNN a model built disregarding domain knowledge on team performance, this architecture overlooks several of the effects outlined in Section~\ref{sec:introduction} as we will show in Section~\ref{sec:performance}. Within this view, to the best of our knowledge, our work represents the only Graph Neural Network model explicitly built to learn disentangled team representations aggregated through attention mechanisms.

\section{The model}

This section introduces the proposed model, \textit{MENTOR}, built by leveraging and extending recent graph representation learning techniques~\cite{GIN, GAT, PGNN}. The model features three main components, which are then aggregated through a soft-attention mechanism \cite{globalattention} that provides expressivity and some degree of interpretability. The final outcome of the model aims to capture the performances reached by teams when living in a graph scenario. In this case, teams are represented by subgraphs and the whole graph represents the ecosystem in which they work. 

\vspace{3mm}
\noindent
In the following sections, we will use the terms \textit{subgraph} and \textit{team} interchangeably.

\subsection{Target definition}

We formally address the problem of modeling team performance as a classification problem. More precisely, we focus on a notion of team performance strongly related to the observed scenario. The most prominent information regarding the teams' outcome, e.g. revenue, public success, ranking position, etc., is summarized in three performance classes, $c_i$: low, middle, and high. We remark how this partition is the result of quantiles of ranking variables that make the three classes ordered, unusually to what happens in a common classification task.

\subsection{Problem formulation}

Let $G =(V,E)$ denote a graph where $V$ and $E$ represent respectively the set of nodes and edges. Each node can be characterized by a set of features $\bm{x}_i \in \mathbb{R}^l$,$i=1,...,|V|$, where $|V|$ is the number of nodes and $l$ denotes the dimensionality of a node's original attributes. As detailed below, we will focus on directed graphs (and hence undirected graphs can easily be recovered as a special case). Moreover, let $S_i=(V_{S_i},E_{S_i})$ be a subgraph of $G$ (i.e., $V_{S_i} \subseteq V $ and $E_{S_i} \subseteq E$) endowed with a discrete label $y_{S_i}$. 
Let $\mathcal{S}=\{S_1,S_2,...,S_n\}$ be a set of subgraphs of interest, our framework allows to model scenarios in which elements of $\mathcal{S}$ may have overlapping nodes. More formally, given $S_i=(V_{S_i},E_{S_i}),S_j=(V_{S_j},E_{S_j}) \in \mathcal{S}$, we could have that $V_{S_i} \cap V_{S_j} \neq \emptyset$. In addition, subgraphs may contain nodes which are not connected to other nodes in the same subgraph.
In other words, some nodes may belong to a team while being completely disconnected from others team members (i.e.: subgraphs can have more than one component). In addition, let us observe that some nodes may not belong to any team, i.e.: $v_k \notin V_i$ where $i=1,...,n$.

Given $\mathcal{S}$, we aim at designing a framework able to generate a $d-$dimensional embedding vector $\bm{z}_{S_i} \in \mathbb{R}^d$ for each $S_i \in \mathcal{S}$ by training a supervised neural model. The final layer of the proposed model consists of a classifier $f: \mathcal{S} \rightarrow \{1,2,...,C\}$ mapping each subgraph $S_i$ to an inferred label $f(S_i)=\hat{y}_{S_i}$. 

\subsection{Proposed model}
\label{sec:proposed_model}

\begin{figure*}[ht]
  \includegraphics[width=\textwidth]{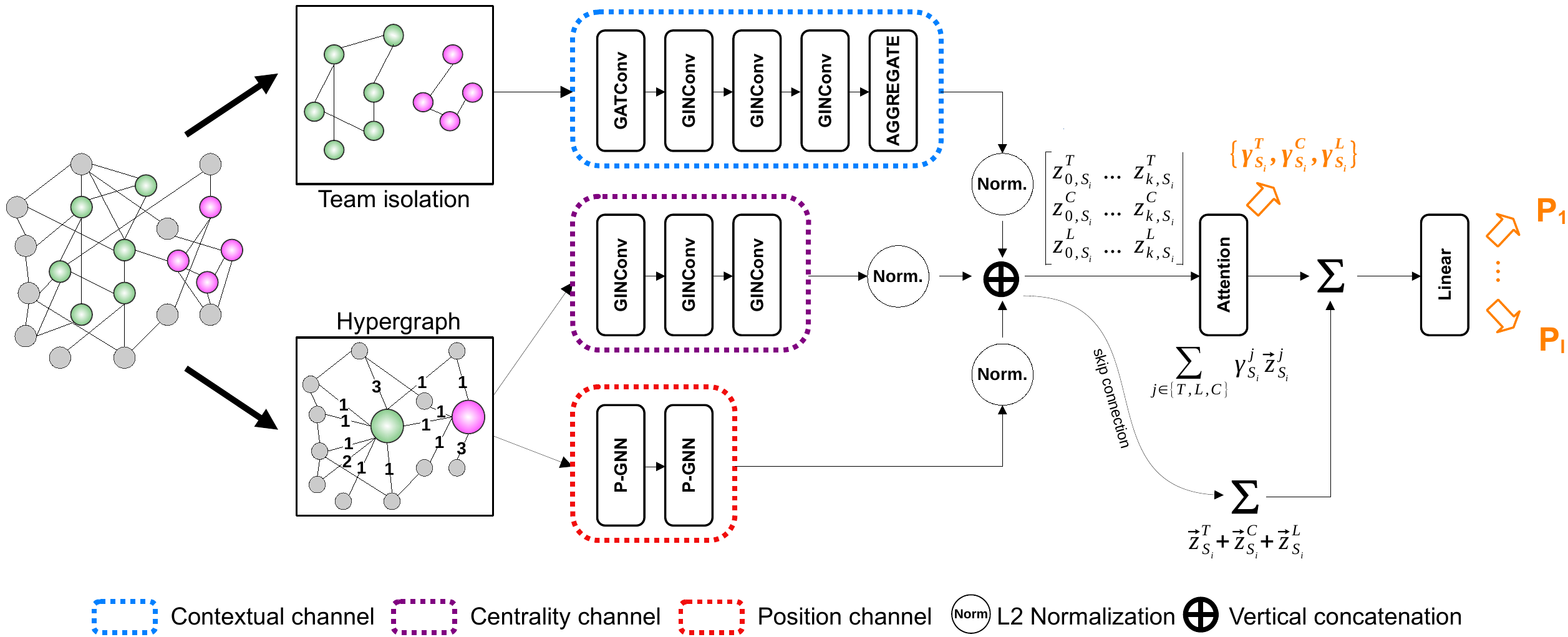}
  \caption{\textbf{MENTOR.} The architecture of our model is based on the usage of three channels: topology (T), centrality (C) and contextual (L). Each channel returns a corresponding embedding vector for each subgraph $S_i$. The outputs of the three channels are then merged by means of an attention mechanism that estimates the importance of a specific effect.}
  \label{fig:Framework}
\end{figure*}

Figure~\ref{fig:Framework} is the overview of our proposed framework. We design a three-channel architecture capable of modeling \textbf{t}opological $(T)$, contextua\textbf{l} $(L)$, and \textbf{c}entrality $(C)$ effects introduced in Section~\ref{sec:introduction}. Each subgraph $S_i$ is independently processed by distinct channels in order to extract different subgraph representations and map $S_i$ to an embedding space: $[\bm{z}_{S_i}^T || \bm{z}_{S_i}^L || \bm{z}_{S_i}^C] \in \mathbb{R}^{3d}$. Downstream, a soft-attention mechanism \cite{globalattention,allyouneed} merges the three components through the estimation of their contribution (in terms of a probability distribution, i.e., $\{\gamma_T, \gamma_L, \gamma_C\}$) to the supervised representation of the subgraph. Analitically, the three channels are merged as follows\footnote{A skip connection between the three-channel representations and the post-attention representation guarantees more stability to the model.}: 

\begin{equation}
    \bm{z}_{S_i} = \gamma_T \bm{z}_{S_i}^{T} + \gamma_L \bm{z}_{S_i}^L + \gamma_C \bm{z}_{S_i}^C, \ \ \ \ \ \ \forall S_i \in \mathcal{S} 
\label{eqn:att}
\end{equation}

\noindent
where $\sum_{i=\{T, L, C \}} \gamma_i = 1$. Concluding, the last layer of the architectures computes labels probabilities, i.e., $f(\bm{z}_{S_i}) = [c_1,c_2,..., c_l]$.

This framework allows us to obtain an expressive model that captures a vast spectrum of network effects. We remark how each channel features a \textbf{preprocessing} phase where the input is parsed into specialized data structures. Besides, a \textbf{computation} phase learns a mapping function to embed arbitrary subgraphs structures into continuous vector representations.

Inspecting equation~\ref{eqn:att}, we observe how the formulation of our model enforces an additive structure of the different channels giving straightforward interpretability on how different effects compose. 

Concluding, we highlight how most of the experiments in GNNs literature apply graph convolutional layers to undirected graphs~\cite{GCN,GraphSAGE,GAT,SUBGNN}. However, in team performance applications (and more in general in social research) is common to encounter scenarios where the direction of the edges conveys crucial information. Therefore,  while building the model, we made sure to inform the message-passing procedures with respect to the directionality of edges by allowing to set different graph convolution directions. For more detail see Appendix~\ref{app:direction_convolution} 

\subsubsection{Topology channel}
\label{sec:topology}

Specific patterns of cooperation may heavily influence team performance. We capture these effects by engineering a branch of the model's architecture that focuses only on interaction patterns captured by the topological structure of each team.

\textbf{Preprocessing} - We design an embedding channel which studies the internal interactions of teams in isolation. In practical terms, we decompose $G$ in a set of non-overlapping $\tilde{S}_i$ subgraphs (i.e., $E_{\tilde{S}_i} \cap E_{\tilde{S}_j} \neq \emptyset$, $\forall i \neq j = 1, ..., n$.) obtained by detaching subgraphs $S_i$, $i=1,...,n$, from the whole graph. Let us remark how this isolation procedure discards all of the nodes not part of a team. Moreover, nodes that belong to multiple teams are replicated into identical disconnected copies in order to obtain non-overlapping subgraphs.

\begin{figure}[ht!]
    \includegraphics[width=0.8\linewidth]{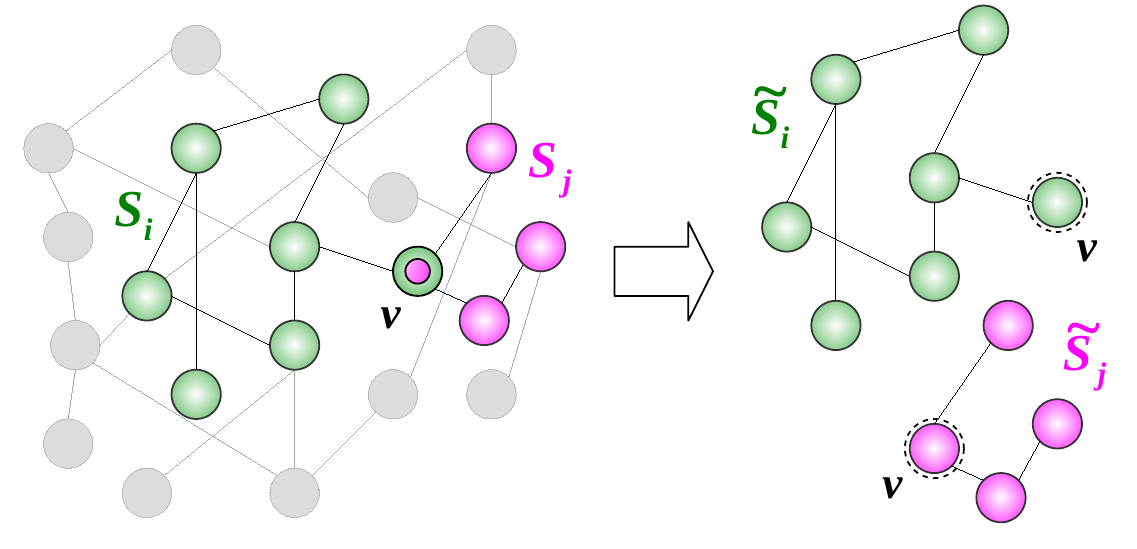}
    \caption{\textbf{Isolation procedure.} Graphical illustration of the isolation procedure of the subgraphs $S_i$ and $S_j$ from $G$, performed by the topology channel. During this phase, the shared member $v$ is duplicated in order to be present in both $\tilde{S}_i$ and $\tilde{S}_j$ subgraphs.}\label{fig:topIso}
\end{figure}

\textbf{Computation} - The isolated subgraphs are mapped to low-dimensional continuous representations exploiting a mixed graph convolutional architecture. Firstly, a single \textit{graph attentional layer}\footnote{We implement this convolutional layer by using GATv2~\cite{GATv2} which fixes several issues of the original GAT layer~\cite{GAT}} (GAT)~\cite{GAT, GATv2} transforms the node features $\bm{X}$ into higher-level representations $\bm{h}^{(1)}$ by pooling information from nodes' 1-hop neighborhood and by learning a self-attention mechanism~\cite{allyouneed,GAT}. Together with embeddings $\bm{h}^{(1)}$, the learned layer returns \textit{attention coefficients}, $\alpha_{vu}$, that indicate the importance of node $u$’s features to node $v$, if they are connected. In more detail:

\begin{gather} 
\bm{h}_v^{(1)} = \alpha_{vv} \bm{W} \bm{x}_v + \sum_{u \in \mathcal{N}(v)} \alpha_{vu} \bm{W} \bm{x}_u \\ 
\alpha_{vu} = \frac{\operatorname{exp}\Big(\bm{a}^\dagger \operatorname{LeakyReLU}(\bm{W}[\bm{x}_v||\bm{x}_u])\Big)}{\sum_{j \in \mathcal{N}(v) \cup \{v\}}\operatorname{exp}\Big(\bm{a}^\dagger\operatorname{LeakyReLU}(\bm{W}[\bm{x}_v||\bm{x}_j])\Big)}
\label{eq:att2}
\end{gather}

\noindent
where $\bm{a} \in \mathbb{R}^{2r}$ and $\bm{W} \in \mathbb{R}^{r \times l}$ are learned quantities, $||$ is the concatenation operator, $\dagger$ represents transposition and $\mathcal{N(\cdot)}$ represents the neighborhood of a given node. The inspection of attention coefficients allows us to understand whether some nodes play a crucial role in the classification task, especially in scenarios where the topological structures may not feature sparse patterns to leverage (Appendix~\ref{app:expressivity} shows an example of the importance of this level of explainability).

Secondly, the next three-layers that complete the topology channel are a modified version of GIN convolution~\cite{GIN}. In more detail, the classical formulation of GIN convolution is extended to accommodate the attention coefficients estimated in the previous layer: 

\begin{equation}
    \bm{h}^{(k)}_v = \theta^{(k)} \Big((1 + \epsilon) \cdot \bm{h}^{(k-1)}_v \cdot \alpha_{vv} + \sum_{u \in N(v)} \bm{h}^{(k-1)}_u \alpha_{vu} \Big)
\label{eqn:GIN}
\end{equation}

\noindent
where $\bm{h}^{(k)}_v$, $k \in \{2,3,4 \}$, is the feature vector of node $v$ at the $k$-th iteration/layer and $\theta$ represents a feed-forward neural network (i.e. an MLP). After $k$ iterations, we learn a representation of the node $\bm{h}_v^{(k)}$ that captures the structural information within its $k$-hop internal network neighborhood. 

Finally, the nodes' embedding vectors are aggregated at team level:

\begin{equation}
    \bm{z}_{S_i}^T = AGGREGATE\Big(\{\bm{h}_v: v \in V_{S_i}\}\Big)
\end{equation}

\noindent
The choice of the $AGGREGATE$ function can be element-wise max-pooling, mean-pooling or add-pooling.

\vspace{3mm}
\noindent
Recent literature on graph representational learning highlights how GNNs are prone to \textit{oversmoothing} issues, i.e., stacking together many layers of graph convolutions results in low variability and similar node level embeddings~\cite{jumping,li2018deeper,klicpera2018predict,godwin2021very}. In the proposed model we mitigate this problem by performing convolutions on subgraphs in isolation and hence preventing message passing operations to be performed on possibly too wide areas of the graphs. Moreover, since the final goal of the channel is to obtain an embedding at the subgraph level by aggregating node level representations, oversmoothing at the node level is not a crucial issue.

\subsubsection{Centrality channel}
\label{sec:centrality}

We capture centrality effects by considering each team as a single entity. We model a team's interactions with the external environment by looking at the links that each team has with others in the ecosystem where it belongs. 

\textbf{Preprocessing} - We collapse each team $S_i$ into a single hypernode which connectivity structure is obtained by rearranging and merging inbound and outbound edges of each node $v \in S_i$. We derive a new graph $H = (V', E')$ where the node $v_i' \in V'$ embodies the $i$-th subgraph $S_i$ of the graph $G$. Let us highlight how edges $e_{ij} \in E'$ are endowed with a weight $w_{ij} \in \mathbb{N}$. The weights are defined by counting how many time nodes belonging to subgraph $S_i$ connect to nodes belonging to subgraph $S_j$. Analitically:

$$w_{ij} = |\{ e_{uv} \in E | u \in S_i, v \in S_j \}|$$

\noindent
In this stage, the hypernodes' features are set considering the teams' original sizes. Note that if some node does not belong to any team, it is regarded as a extra hypernode in $H$. 

\begin{figure}[ht!]
    \includegraphics[width=0.8\linewidth]{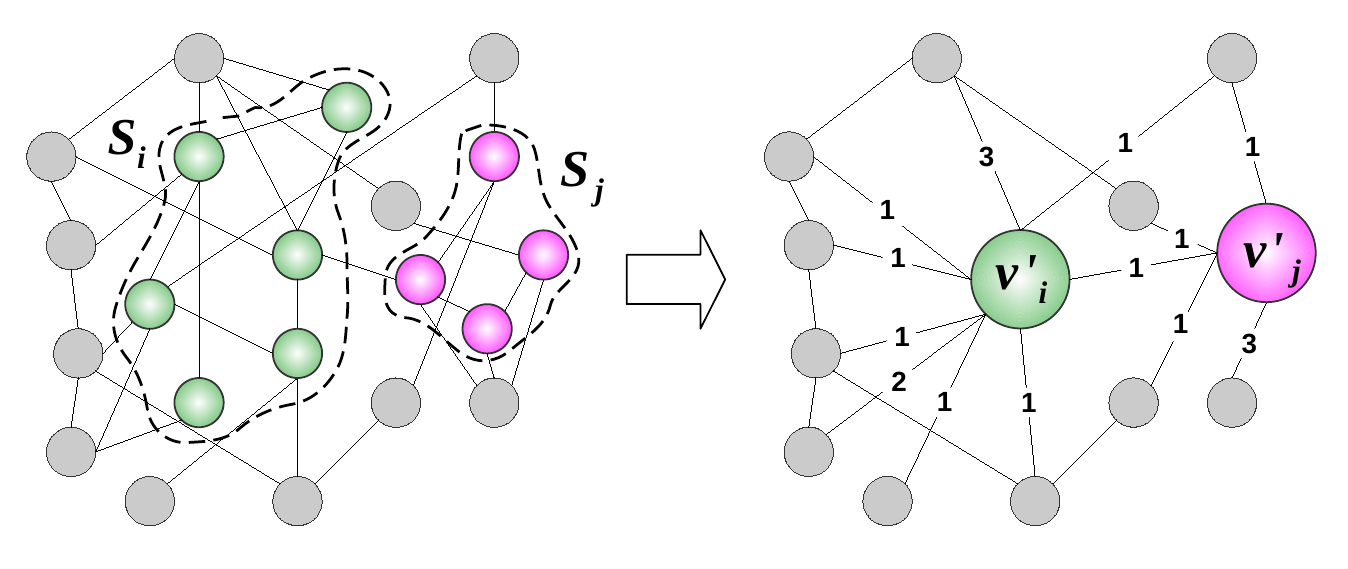}
    \caption{\textbf{Hypernodes creation.} Graphical illustration of the preprocessing phase of the centrality channel: subgraphs $S_i$ and $S_j$ of $G$ are collapsed to the hypernodes $v_i'$ and $v_j'$. Edges in the new hypergraph are weighted according to the connectivity structure of the original graph $G$.}\label{fig:b}
\end{figure}

\textbf{Computation} - As regards the architectural aspect, we employ the modified version of GIN illustrated in equation~\ref{eqn:GIN}. In more detail, we model the structural information of a $3$-hop weighted network neighborhood by means of three convolutional layers, where the attention coefficients $\alpha$ are replaced with the current weights $w$. These hypernode-level iterations deliver embeddings $\bm{z}_{S_i}^C$ related to subgraph $S_i$. 

\subsubsection{Contextual channel}

Contextual effects in a graph structured environment tell us that nodes at close distance (in terms of hops) likely feature similar underlying characteristics. Also, in this case, we consider teams as a single entity assuming that members inside the team feature a zero distance.

\textbf{Preprocessing} - In this channel, we exploit the formulation of the hypergraph $H$ introduced in the preprocessing chapter of Section~\ref{sec:centrality}. In more detail,  hypergraph $H$ is populated by the \textit{hypernode} teams and our goal is to obtain the contextual embeddings $\bm{z}_{S_i}^L$, $\forall v_i' \in V' | S_i \in \mathcal{S}$.

\begin{figure}[ht!]
    \centering
    \includegraphics[width=0.5\linewidth]{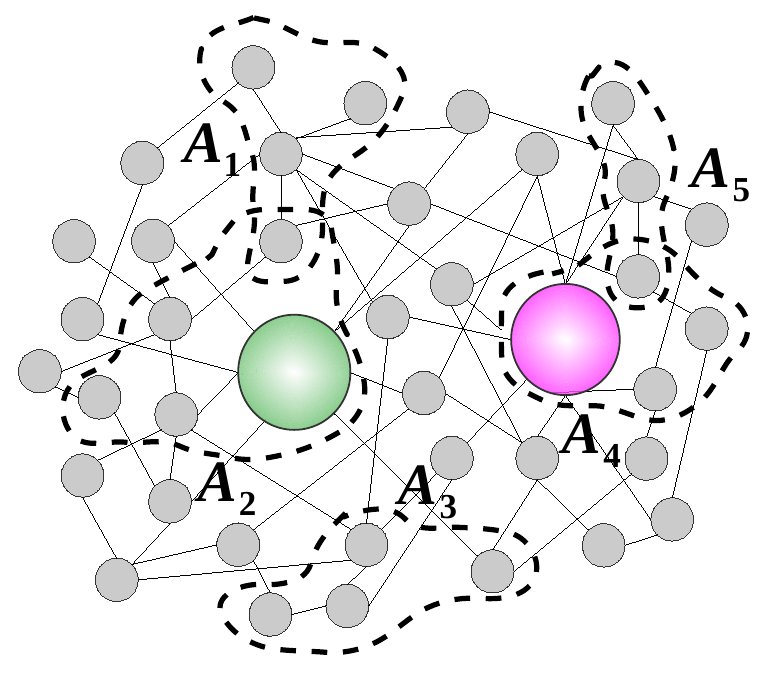}
    \caption{\textbf{Anchor-sets.} Graphical illustration of the anchor-sets $A_i$ generated by the P-GNN algorithm in order to potentially cover the entire volume of the graph $H$.}\label{fig:anchor}
\end{figure}

\textbf{Computation} - A drawback of recently developed graph convolutional architectures~\cite{GraphSAGE,GAT,GIN} is their inability to model contextual traits of nodes in the broader context of the graph structure~\cite{PGNN,srinivasan2019equivalence}. For example, suppose two nodes belong to different areas of the network (i.e., they are many hops apart with respect to the graph diameter) but have topologically the same (local) neighborhood structure. In that case, they will have identical embedding representations~\cite{PGNN,srinivasan2019equivalence}. For this reason, we decide to exploit the P-GNN \cite{PGNN} approach for computing position-aware node embeddings. The standard convolutional methods aggregate features from the node’s local network neighborhood while P-GNN involves using some \textit{anchor-sets} $A_i$, subsets of nodes of the graph (see Figure~\ref{fig:anchor}), as reference points from which to learn a non-linear distance-weighted aggregation
scheme. 

\noindent
By exploiting this convolutional layer we encode the global network position of a given node. More precisely, P-GNN returns an embedding vector $\bm{z}_{S_i}^L \in \mathbb{R}^s$, where $s$ is the number of anchor-sets. Through a linear transformation, we then adapt the contextual embedding to be $d$-dimensional. Moreover, we decide to learn contextual representations without considering nodes' attributes. We remark how the regular P-GNN architecture requires computing the shortest path matrix of the modeled graph. As soon as the network grows in the number of nodes to be modeled, the computational requirements of this method (even with the proposed approximated version) scale quadratically. Structuring the input as a hypergraph, as proposed earlier, helps in mitigating such computational requirements by greatly reducing the number of nodes and therefore also the number of shortest paths to be computed. Concluding, the architecture features two layers of P-GNN.

\subsubsection{Aggregation mechanism}

Before feeding the embedding delivered by the three channels into the aggregation mechanism, each $\bm{z}_{S_i}^j$ is normalized as follows:

\begin{equation}
    \bm{z}_{S_i}^j = \frac{\bm{z}_{S_i}^j}{max(||\bm{z}_{S_i}^j||_2, \epsilon)}, \ \ \ \ \ j \in \{T, P, C\}
\end{equation}

According to the equation~\ref{eqn:att}, we insert an attention mechanism that boosts model expressivity while quantitatively estimating how different effects compose. The three output embeddings are then merged in order to estimate the importance of a specific effect conditioned on 1) the single observation 2) the modeled dataset. 

The final embedding is then obtained by a soft-attention mechanism inspired by Yujia Li et al.~\cite{globalattention}:

\begin{equation}
    \bm{z}_{S_i} = \sum_{j \in \{T,C,L\}} \gamma_{S_i}^j \bm{z}_{S_i}^j
\end{equation}

\noindent
where $\gamma_{S_i}^j$ is the attention coefficient and is computed as:

\begin{equation}
    \gamma_{S_i}^j = \frac{e^{\theta_{gate}(\bm{z}_{S_i}^j)}}{\sum_{k \in \{T,C,L\}} e^{\theta_{gate}(\bm{z}_{S_i}^k)}}
\label{eq:attchannel}
\end{equation}

\noindent
where $\theta_{gate}$ represents a 2-layer MLP.

\section{Data}

To assess our framework’s capabilities, we perform extensive experiments on both synthetic and real-world datasets. Our work focuses on modeling team performance; however, it is important to stress how a clear-cut notion of team performance is not always identifiable and may be an object of debate. Moreover, given the heterogeneity of the scenarios that we address, encoding the problem into the graph structure may be context-dependent. Therefore, in order to obtain the datasets listed below, we formulate several working hypotheses followed by different pre-processing steps. 

\subsection{Synthetic datasets}

We create several synthetic datasets where team performance is appropriately encoded to reflect different effects. In particular, we engineered scenarios according to the factors introduced in Section~\ref{sec:introduction}. All of the synthetic datasets do not feature attributes at the node level (that are therefore initialized as a constant for all nodes). Crucially, these tasks are designed to assess the capabilities of different models to leverage connectivity structures of graphs and subgraphs.

\vspace{3mm}
\noindent
We report details, visualizations and algorithmic procedures used to generate and analyze the datasets listed below in Appendix~\ref{app:data}. The code to reproduce these datasets is released together with the proposed model. 

\vspace{3mm}
\textbf{Centrality [CIn} \& \textbf{COut]} - We create two different directed graphs of $10\,008$ nodes populated by 990 teams where the teams' label is driven respectively by the value of the in-degree (graph with $64\,480$ edges) and out-degree (graph with $64\,477$ edges) of one team respect to others. All of the teams share a common internal connectivity structure (i.e., generated by the same model) which is superimposed onto a random graph generated by a (directed) Erd\H{o}s-R\'enyi (ER) model~\cite{erdos1960evolution}. The added connectivity patterns are unrelated to the final teams' targets and can be seen as ways to add noise to the classification problem. We then sort teams into three different groups and aim to obtain differentiated subgraphs by manually adding in (or out) external connectivity patterns. In practical terms, we enrich existing teams' connections by adding extra edges according to the group into which a team is sorted into. Teams in group 0 will receive a small number of extra edges with respect to the initial endowment whereas teams in group 2 will receive more (for more details one can see algorithm \ref{alg:CIn} and figure \ref{fig:Centrality} in appendix \ref{app:centrality}). 

\vspace{3mm}
\textbf{Topology [T1]} - A directed graph of $10\,008$ nodes and $70\,699$ edges. The nodes are partitioned into 990 variable size subgraphs where the connectivity pattern between their own members is generated by repeatedly adding three predefined motifs (delivering a 3-class problem, see figure \ref{fig:Motifs} in appendix \ref{app:Topology}). This dynamic is superimposed onto a random graph generated by a (directed) Erd\H{o}s-R\'enyi (ER) model. The labels in this dataset are determined by the type of motif used to define the internal team structure. The parameter $m$ of the underlying connectivity defined by the ER model allows us to control the amount of global noise by adding random edges not relevant to the classification problem (see Appendix~\ref{app:Topology} and algorithm \ref{alg:T1}). 

\vspace{3mm}
\textbf{Topology [T2]} - A directed graph of $10\,008$ nodes and $64\,518$ edges. The formulation of this dataset follows closely the one exposed in \textbf{[T1]} and \textbf{[CIn} \& \textbf{COut]}. However, \textbf{[T1]} features the connectivity structure introduced in \textbf{[CIn} \& \textbf{COut]} with less stochasticity from the underlying ER base. Nonetheless, the target is still defined by the patterns defined in \textbf{[T1]}. Therefore, with this dataset we asses the robustness of models with respect to structured noise.

\vspace{3mm}
\textbf{Topology [T3]} - A directed graph of $10\,000$ nodes and $70\,721$ edges. Identical to \textbf{[T1]} construction but featuring teams with overlapping members and $1\,670$ nodes not belonging to any team. After having introduced this last Topology dataset we remark how using motifs to build \textbf{T*}s datasets represents a flexible way to encode in synthetic data different possible types of cooperation patterns that may take place inside a team.

\vspace{3mm}
\textbf{Contextual [L]} - A Caveman graph \cite{caveman} populated by $2\,000$ nodes and $9\,000$ edges. Cliques of the caveman represent our teams. Labels of the teams are given by trisecting the ring structure of the graph and assigning each group to a label.
 
\vspace{3mm}
\textbf{Contextual\&Topology [LT]} - A directed graph of $9\,956$ nodes and $28\,504$ edges. The nodes are partitioned into $1\,000$ teams with stochastically sampled sizes (see algorithm \ref{alg:team_maker} for more details). The labels in this dataset are determined by a combination of \textit{contextual} and internal \textit{topological} effects. We created a scoring system that allows us to combine different metrics. \textit{Contextual effects} are determined by endowing each team with a pair of polar latent coordinates which map them into a 2-dimensional Euclidean space. Radius and angle respectively determine the magnitude and sign of the contextual score. \textit{Topological effects} are determined by computing the Gini index of the in-degree distribution of the nodes inside a subgraph. Teams' internal connectivity is generated according to two different patterns (associated with low and high Gini) and then randomly rewired. For more details refer to appendix \ref{app:position_topology}.

\subsection{Real-world datasets}

We also study real-world datasets spanning a spectrum of contexts going from casts of movies to data scientists working together towards the solution of a predictive task. In contrast with what illustrated in synthetic data, real-world datasets feature a plethora of nodes attributes and the final target may not be solely a function of the graph connectivity structure. We highlight how the raw data was pre-processed and we provide more details in this regard in Appendix~\ref{app:Real datasets}.

\vspace{3mm}
\textbf{IMDb} - An undirected graph of $4\,802$ nodes and $25\,632$ edges. The graph features $586$ teams (that are represented by films). In more detail, this dataset is obtained by subsampling the full IMDb dataset and considering only films after 2018 (included). The connectivity structure of the graph encodes cast (actor/actress, director, producer, composer, etc.) co-working in different films. The cast of the movie is represented as a clique and cast components working in more than one film serve as bridges in the graph connecting different cliques. Targets in this dataset are defined by discretizing into three classes (by means of quantiles) the absolute income of films released in a predefined time window. 

\vspace{3mm}
\textbf{Dribbble} - A directed graph of $5\,196$ nodes and $304\,315$ edges. The graph features $769$ possibly overlapping teams. The connectivity structure of the graph encodes the "follow" interactions featured in a static snapshot of the Dribbble.com social network (nodes as users, edges as follow interactions). Dribbble is a social platform that allows users to organize themselves in teams with the aim of creating and sharing digital art through so-called \textit{shots} (i.e., posts). Labels are determined by discretizing into three classes (by means of quantiles) the number of likes received by creative content in a predefined time window.

\vspace{3mm}
\textbf{Kaggle} - A directed graph of $4\,183$ nodes and $17\,789$ edges. The nodes are partitioned into $1\,013$ variable size overlapping subgraphs and the global connectivity structure is built on the basis of a static snapshot of Kaggle.com "follow" network. Moreover, being the "follow" network poorly populated, we add an extra connectivity structure based on co-working, similar to IMDb. Kaggle is a competition platform for predictive modeling where individual users or teams can participate to solve a task and be consequently ranked relative to the others. Targets in this dataset are defined by discretizing into three classes (by means of quantiles) the average ranking position of the teams in a predefined time window. 

\section{Experiments}

\begin{table*}[ht]
    \centering
    \begin{tabular}{m{1em} c c c c c c c c}
        &\\
        \hline
        \rotatebox{90}{\textbf{Datasets}}& \makecell{Dataset \\ \#teams\\ \#classes \\ \#nodes \\ \#edges \\ Class distr.} & \makecell{CIn \\ $990$ \\ $3$ \\ $10 \,008$ \\ $64\,480$ \\ $33/33/33$} & \makecell{COut \\ $990$ \\ $3$ \\$10\,008$ \\ $64\,477$ \\ $33/33/33$} & \makecell{T1 \\ $990$ \\ $3$ \\ $10\,008$ \\ $70\,699$ \\ $33/33/33$} & \makecell{T2 \\ $990$ \\ $3$ \\ $10\,008$ \\ $64\,518$ \\ $33/33/33$}& \makecell{T3 \\ $1\,000$ \\ $3$ \\ $10\,000$ \\ $70\,721$ \\ $33/33/33$} & \makecell{L \\ $200$ \\ $3$ \\ $2\,000$ \\ $9\,000$ \\ $33/33/33$}& \makecell{LT \\ $1\,000$ \\ $3$ \\ $9\,956$ \\ $28\,504$ \\ $33/33/33$} \\
        \hline
        \rotatebox{90}{\textbf{Classical ML}}& \makecell{LR \\ SVM\\ RF \\ XGBoost \\ MLP} & \makecell{\bm{$99.8 \pm 0.3$} \\ \bm{$99.8 \pm 0.3$} \\ $99.5 \pm 0.5$ \\ $99.6 \pm 0.5$ \\ $99.0\pm 0.0$} & \makecell{\bm{$100.0 \pm 0.0$} \\ \bm{$100.0 \pm 0.0$} \\ \bm{$100.0 \pm 0.0$} \\ \bm{$100.0 \pm 0.0$} \\ $100.0\pm 0.1$} & \makecell{$43.3 \pm 3.3$ \\ $43.0 \pm 3.5$ \\ $45.8 \pm 2.4$ \\ $45.4 \pm 2.6$ \\ $44.8\pm2.0$} & \makecell{$46.3 \pm 2.8$ \\ $46.8 \pm 4.3$ \\ $47.0 \pm 4.7$ \\ $47.0 \pm 4.8$ \\ $47.3 \pm 4.3$}& \makecell{$48.7 \pm 2.2$ \\ $50.5 \pm 4.6$ \\ $48.0 \pm 4.1$ \\ $47.7 \pm 3.9$ \\ $51.1 \pm 3.4$} & \makecell{$33.8 \pm 1.3$ \\ $35.0 \pm 0.0$ \\ $33.0 \pm 1.1$ \\ $32.5 \pm 0.0$ \\ $33.0 \pm 1.1$}& \makecell{$58.4 \pm 2.1$ \\ $60.1 \pm 1.9$ \\ $59.2 \pm 2.6$ \\ $58.2 \pm 2.5$ \\ $60.3 \pm 1.6$} \\
        \hline
        \rotatebox{90}{\textbf{GNN}}& \makecell{SubGNN \\ MENTOR - T \\ MENTOR - C \\ MENTOR - L \\ MENTOR} & \makecell{$57.5 \pm 15.2 $ \\ $33.6 \pm 3.8$ \\ $99.5 \pm 0.4$ \\ $62.1 \pm 2.2$ \\ $99.7 \pm 0.4$} & \makecell{$ 62.1 \pm 22.8 $ \\ $34.4 \pm 2.7$ \\ \bm{$100.0 \pm 0.0$} \\ $77.2 \pm 14.8$ \\ \bm{$100.0 \pm 0.0$}} & \makecell{$ 37.6 \pm 6.1 $ \\ \bm{$96.2 \pm 1.2$} \\ $31.7 \pm 3.5$ \\ $34.6 \pm 3.9$ \\ $95.2 \pm 1.4$} & \makecell{$ 35.6 \pm 8.9 $ \\ \bm{$95.6 \pm 1.7$} \\ $33.7 \pm 2.6$ \\ $34.3 \pm 4.2$ \\ $95.4 \pm 1.8$}& \makecell{$49.3 \pm 8.0 $ \\ $96.0 \pm 0.6$ \\ $39.9 \pm 5.4$ \\ $32.9 \pm 2.3$ \\ \bm{$96.6 \pm 1.3$}} & \makecell{$94.4 \pm 1.9 $ \\ $32.5 \pm 0.0$ \\ $33.0 \pm 1.1$ \\ \bm{$99.3 \pm 1.2$} \\ \bm{$99.3 \pm 1.2$}}& \makecell{$ 62.2 \pm 7.1 $ \\ $60.5 \pm 1.9$ \\ $43.4 \pm 3.0$ \\ $53.3 \pm 4.4$ \\ \bm{$87.2 \pm 2.2$}} \\
        \hline
    \end{tabular}
    \caption{\textbf{Accuracy on synthetic datasets}. Standard deviations are provided from runs with 10 random seeds.}
    \label{tab:results1}
\end{table*}


\subsection{Learning setup}
\label{sec:learning_setup}

For all the considered datasets, we apply a train/test split on teams labels using a ratio of 80/20. On each model run, we perform a bayesian hyper-parameter search procedure~\cite{tpe,tpe2} by evaluating the validation performance of the model through a 5-fold validation using the Optuna optimization library~\cite{optuna} (monitoring the val loss as optimization objective function). The space of hyper-parameter we swept is quite large and detailed information about the procedure can be found in Appendix~\ref{app:implementation}. To assess the stability of the proposed architecture with respect to various tasks, we kept architectural hyper-parameters (i.e., the number of convolutional layers in each channel) fixed. This allowed us to gauge the performance of our model "out of the box" with combinations of hyper-parameters that may be sub-optimal. 

\noindent
The model is trained using the Adam optimizer~\cite{adam}. Moreover, in order to achieve better generalization and more stable results we make use of the Stochastic Weight Averaging ensembling technique~\cite{swa}. During the development of the proposed architectures, we encountered several instability issues related to the training procedure. We fixed such issues by adding to the model a skip layer~\cite{resnet} (see Figure~\ref{fig:Framework}).

\subsection{Baselines definition}
\label{sec:compared}

Let us now introduce the baselines with respect to which the proposed model is compared. In order to achieve a thorough assessment of the architecture's performance, we test the model against several classical machine learning algorithms. In more depth, we compare the proposed model against logistic regression (LR), support vector machines (SVM), random forests \cite{breiman2001random} (RF), boosting methods \cite{chen2016xgboost} (XGBoost) and multi-layer perceptron (MLP). As for the graph neural network side, we test the SubGNN model which is the most significant contribution able to handle subgraph structures. Moreover, we point out how the single channels of our model can each serve as baseline (on the basis of popular graph neural network algorithms, more details in Section~\ref{sec:Ablation}).

\noindent
Let us remark how, in classical machine learning methods, feature engineering and aggregation function specification need to be defined. Classical ML algorithms heavily rely on tabular data and therefore on domain knowledge and hand-engineered features. When working with subgraphs, several features are defined directly on the graph structure as a whole whereas other features are defined at the node-level. In particular, for node-level features, an aggregation function (i.e., max, min, mean, sum) needs to be specified to obtain the required subgraph representation in terms of tabular format. The list of features used for these models is reported in Appendix~\ref{app:empirical_eval}.

\subsection{Performance comparison}
\label{sec:performance}

We evaluate and compare the test performances of the models following the learning instructions explained in Section~\ref{sec:learning_setup}. In particular, we run each model by defining 10 different random seeds and hence obtaining different train/validation/test splits. This setting allows us to test the generalization power and the robustness with respect to the randomness of the various methods. We will report the performances of the proposed model and the relative baselines in terms of accuracy. Moreover, we will report in Appendix~\ref{app:empirical_eval} models' performance with respect to AUROC metric.

\begin{figure*}[h!]   
\centering
\subfloat[IMDb]{\includegraphics[width=0.33\textwidth]{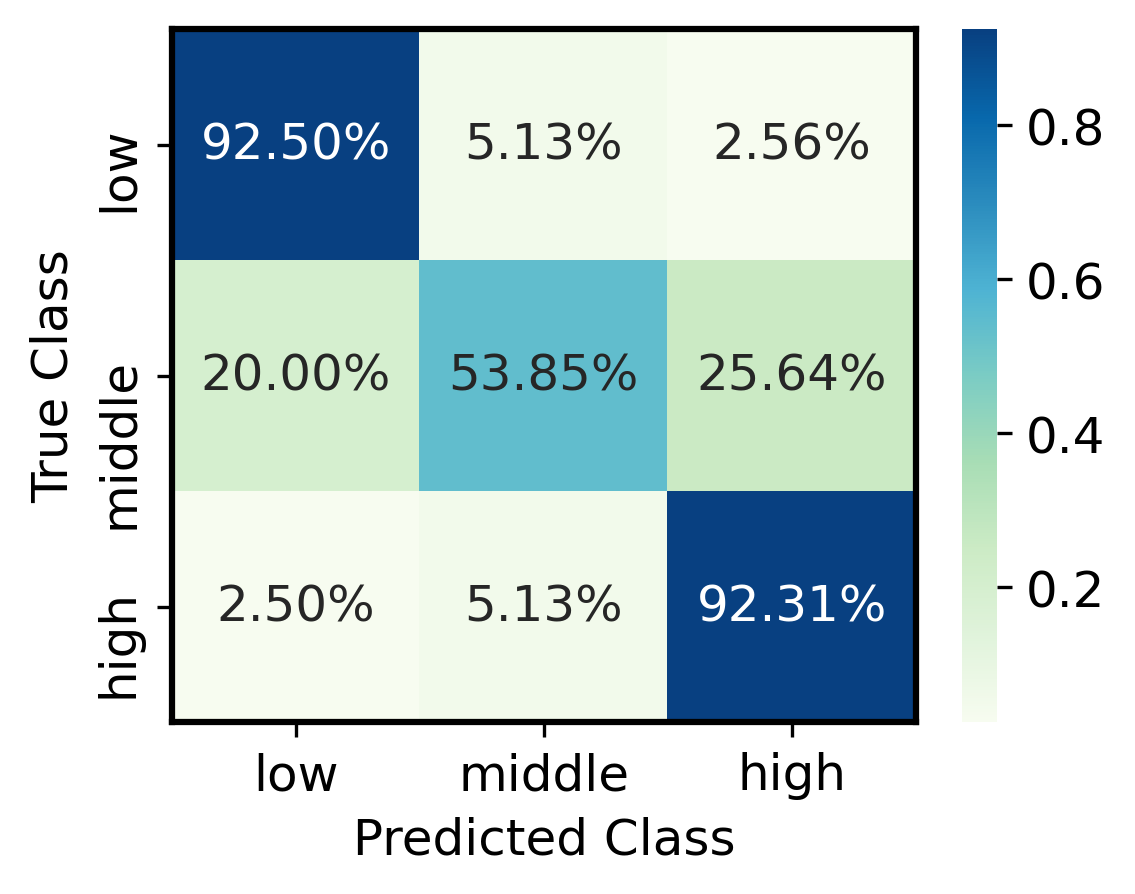}}
\subfloat[Dribbble]{\includegraphics[width=0.33\textwidth]{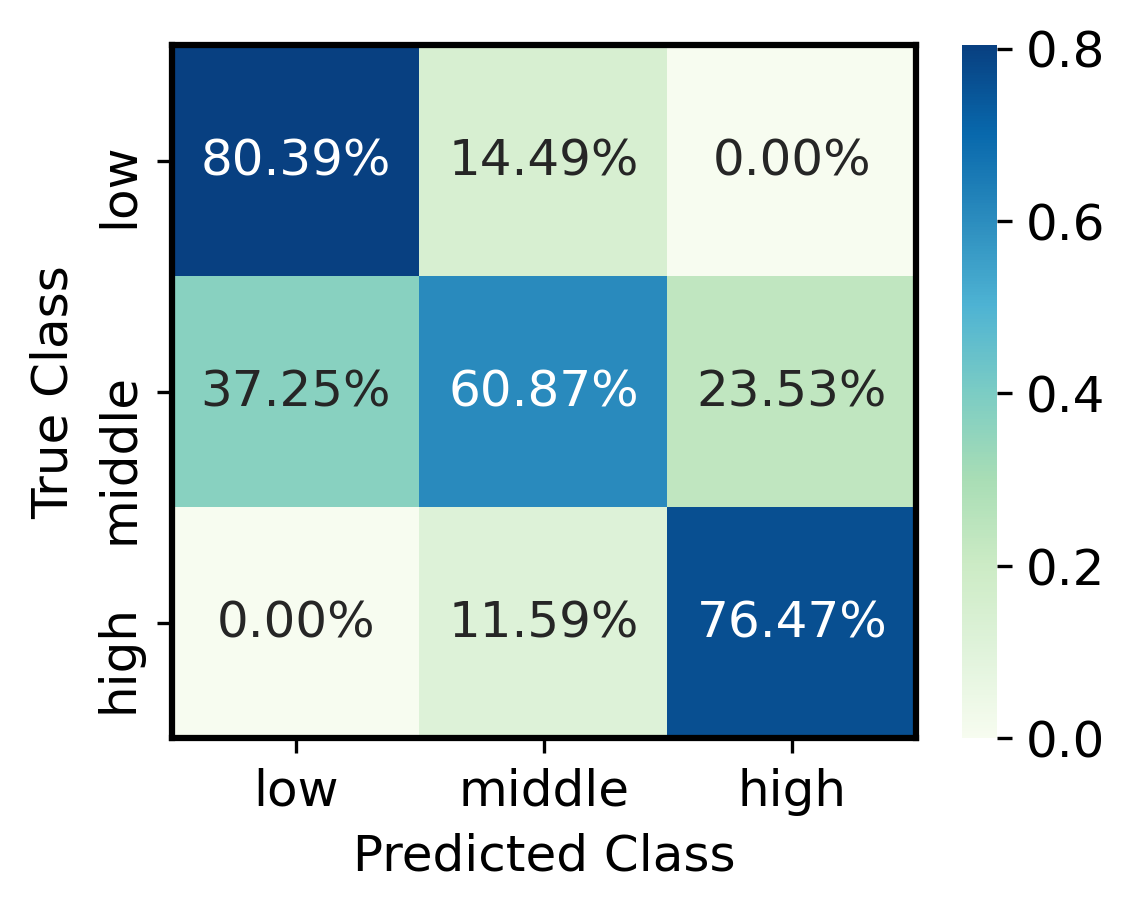}}
\subfloat[Kaggle]{\includegraphics[width=0.33\textwidth]{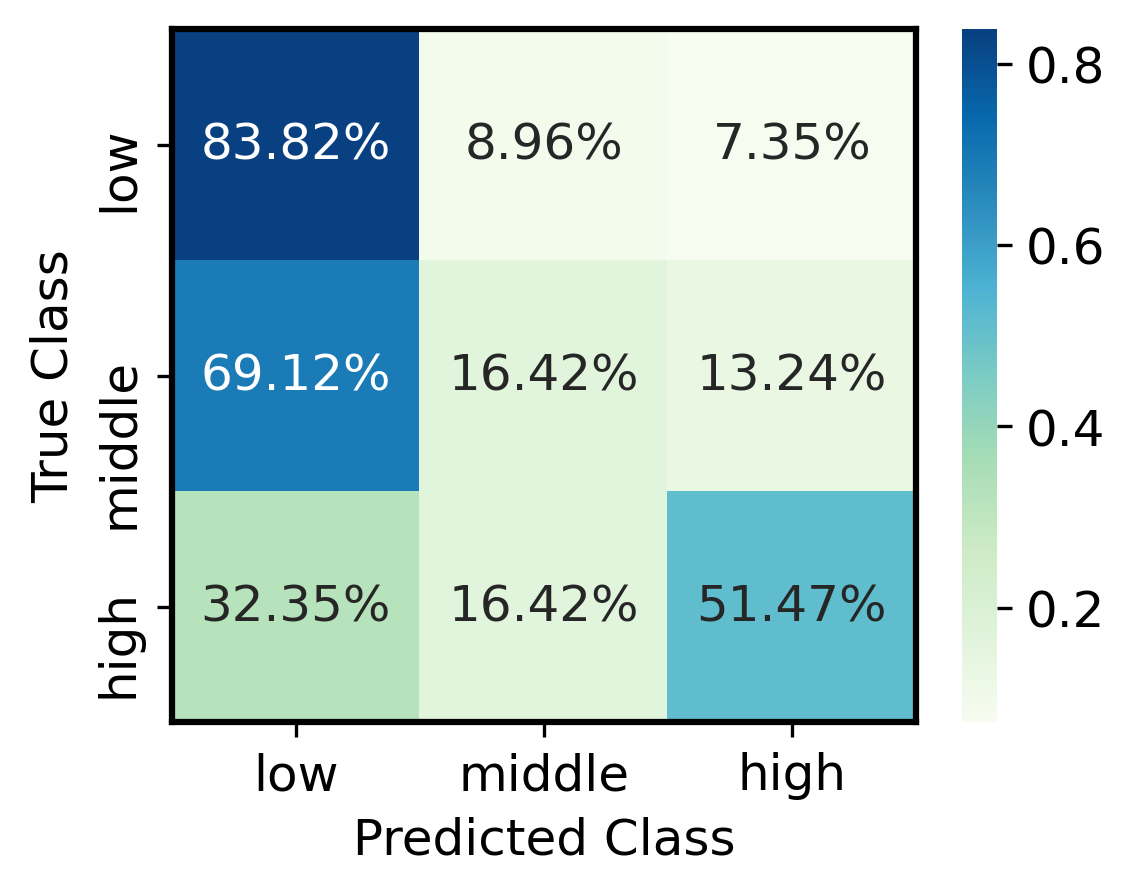}}

\caption[Optional caption for list of figures 5-8]{\textbf{Confusion matrices on real-world datasets.} The confusion matrices on real-world datasets: a) IMDb; b) Dribbble c) Kaggle. The results refer to the configuration corresponding to the seed which returns the highest accuracy.}
\label{fig:real_cf}
\end{figure*}

\vspace{3mm}
\noindent
\textbf{Synthetic datasets.} Results are shown in Table~\ref{tab:results1}. On the one hand, performances on the centrality datasets (CIn, COut) are excellent for all of the tested models except for SubGNN. On the other hand, results on the topology datasets (T1, T2, T3) show how the proposed framework significantly outperforms all baselines by $52\%$ on average. Results on the contextual dataset (L) show how SubGNN and MENTOR vastly outperform classical baselines. However, we highlight how MENTOR's accuracy exceeds SubGNN's by $4.8\%$. Concluding, performances on the mixed-effects dataset (PT) show how our model largely beats all the other baselines by $27.8\%$ on average.

\noindent

\begin{table}[ht]
    \centering
    \begin{tabular}{m{1em} c c c c c}
        &\\
        \hline
        \rotatebox{90}{\textbf{Datasets}}& \makecell{Dataset \\ \#teams\\ \#classes \\ \#nodes \\ \#edges \\ Class distr.} & \makecell{IMDb \\ $586$ \\ $3$ \\ $4\,802$ \\ $25\,632$ \\ $33/33/33$} & \makecell{Dribbble \\ $769$ \\ $3$ \\ $5\,196$ \\ $304\,315$ \\ $33/45/22$} & \makecell{Kaggle \\ $1\,013$ \\ $3$ \\ $4\,183$ \\ $17\,789$ \\ $33/33/33$}\\
        \hline
        \rotatebox{90}{\textbf{Classical ML}}& \makecell{LR \\ SVM\\ RF \\ XGBoost \\ MLP} & \makecell{$63.8 \pm 4.8$ \\ $64.4 \pm 4.9$ \\ $63.4 \pm 4.5$ \\ $64.2 \pm 4.3$ \\ $  64.1 \pm 4.0$} & \makecell{$62.5 \pm 2.4$ \\ $63.8 \pm 1.8$ \\ $64.4 \pm 4.0$ \\ $64.2 \pm 3.5$ \\ $64.5 \pm 2.2$} & \makecell{$\bm{47.2 \pm 2.7}$ \\ $46.4 \pm 2.0$ \\ $44.7 \pm 3.2$ \\ $46.3 \pm 2.5$ \\ $47.0 \pm 2.6$}\\
        \hline
        \rotatebox{90}{\textbf{GNN}}& \makecell{SubGNN \\ MENTOR - T \\ MENTOR - C \\ MENTOR - L \\ MENTOR} & \makecell{$63.3 \pm 3.1  $ \\ $66.9 \pm 7.1$ \\ $51.7 \pm 4.6$ \\ $50.0 \pm 3.7$ \\ \bm{$69.1 \pm 5.0$}} & \makecell{$ 62.1 \pm 4.1 $ \\ $63.0 \pm 3.8$ \\ $63.0 \pm 3.8$ \\ $45.2 \pm 2.4$ \\ \bm{$66.0 \pm 3.2$}} & \makecell{$ - $ \\ $45.1 \pm 1.8$ \\ $46.6\pm 2.1$ \\ $45.4 \pm 2.7$ \\ $45.3 \pm 3.3$}\\
        \hline
    \end{tabular}
    \caption{\textbf{Accuracy on real-world datasets}. Standard deviations are provided from runs with 10 random seeds.}
    \label{tab:results2}
\end{table}

\noindent
\textbf{Real-world datasets.} Results are shown in Table~\ref{tab:results2}. On IMDB, the classical machine learning methods show comparable performances while our model outperforms them by $5.1\%$. On Dribbble, the proposed framework outperforms all of the baselines by $2.7\%$ on average. The results on Kaggle show an overall poor performance where the logistic regression, the best model, reaches an accuracy of only $47.2\%$. The results of all the models suggest that the information available about the dataset may not be sufficient or not well-defined in order to solve the current task. We remark how, for the Kaggle dataset, embedding teams by means of SubGNN was not possible. The
 training procedure requires the input graph to be fully connected. This is one of the limitations featured by SubGNN that we address in proposing our MENTOR. 

\noindent
The confusion matrices in Figure~\ref{fig:real_cf} show how the model fails more often in classifying the middle class, which acts as a "bridge" between the boundary labels. Let us remark that this is somewhat expected since labels for real-world datasets are obtained through a discretization by means of quantile partitioning.  Therefore,  classes can be poorly separated at cut-off values by design. On the contrary, the model rarely confuses the high class with low class and vice-versa (it never happens for the Dribble dataset), correctly guessing with high accuracy in the boundary classes (about $92\%$ for IMDb dataset).

\subsection{Ablation study}
\label{sec:Ablation}

We perform ablation studies to understand whether the specific channels of the architecture are capable of capturing the effects which they were designed for. In particular, we compute the model's metrics by turning off all the channels but one and compare the results with the whole architecture. Let us remark on how the single channels of our model can be seen as further benchmarks of the proposed architecture against graph neural network baselines (i.e., GIN~\cite{GIN}, GAT~\cite{GAT}, P-GNN~\cite{PGNN}). As explained in Section~\ref{sec:proposed_model}, the different preprocessing phases redesign the input graph in two main structures: 1) subgraphs in isolation (topology); 2) subgraphs condensed into hypernodes (centrality and contextual). Firstly, this setting allows the topology channel to embed subgraphs by classifying isolated substructures as standalone graphs. Secondly, the centrality and contextual channels leverage node-level learning architectures applied on the pooled original graph (where nodes encode subgraphs). As shown in Table~\ref{tab:results1}, the removal of certain channels can lead to an increase in performances with respect to the 3-channel setting,  however, these percentage increases are not striking ($1\%$ max.). This result is very comforting considering that in a real-world scenario the effects that drive the analyzed system are not known a priori. 

\begin{figure}[htp]
\centering
\subfloat[L]{%
  \includegraphics[clip,width=0.6\columnwidth]{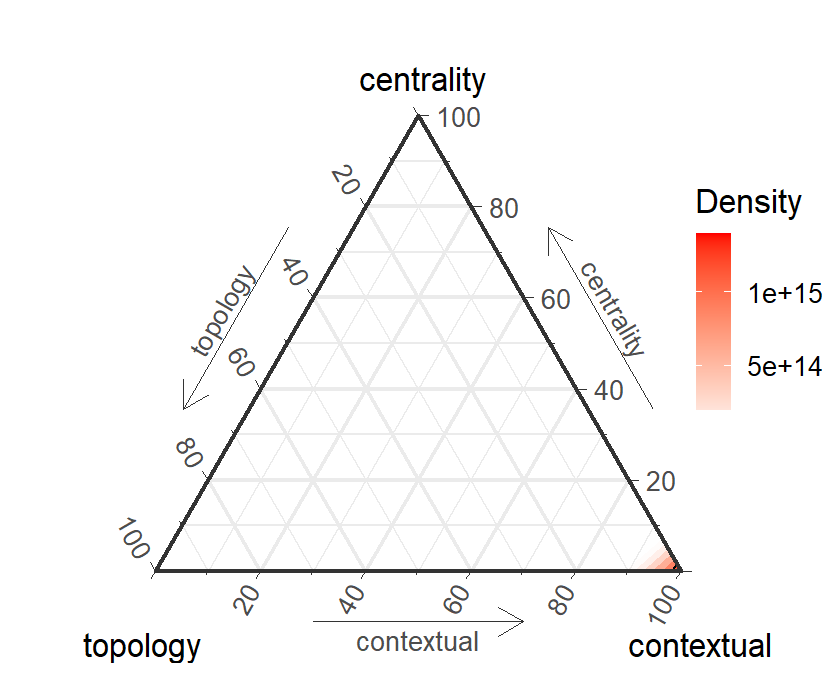}%
}

\vspace{-.20cm}

\subfloat[LT]{%
  \includegraphics[clip,width=0.6\columnwidth]{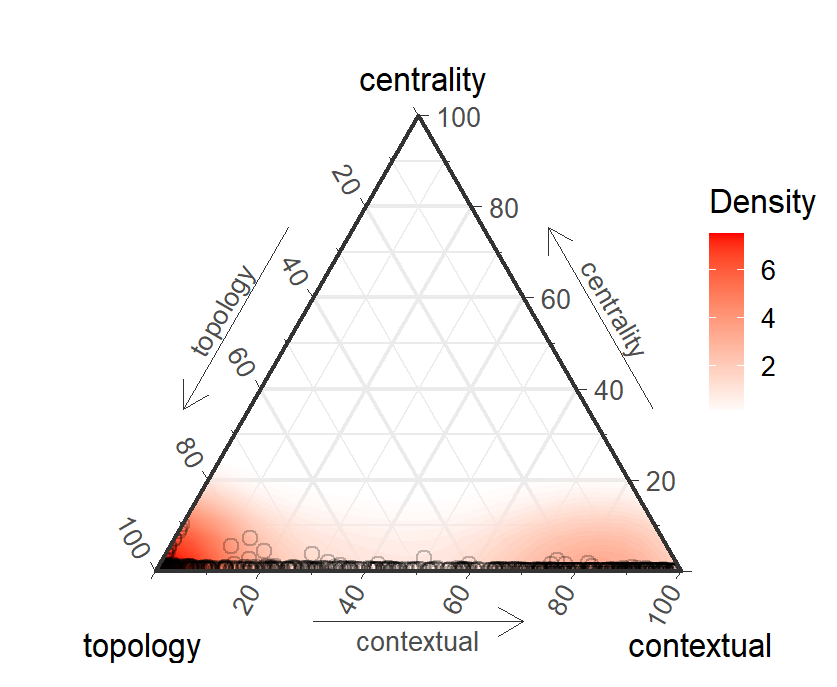}%
}
\caption{\textbf{Ternary graph of the attention coefficients on two synthetic datasets}. The attention coefficients explained in Formula~\ref{eq:attchannel} visualized by means of a ternary graph. The coefficients are obtained by training MENTOR on a) synthetic dataset L and b) synthetic dataset LT. The results refer to the configuration corresponding to the seed which returns the highest accuracy.}
\label{fig:attsy}
\end{figure}


\noindent
Let us remark that since each channel performs well with respect to the effect it is designed to capture while performing poorly in the residual scenarios, single-channel embeddings are likely to be uncorrelated. This feature should boost the-reliability of attention coefficients highlighting non-overlapping contribution of different effects to the final outcome.

\subsection{Model findings}
\label{sec:findings}

\begin{figure*}[ht]   
\centering
\subfloat[IMDb]{\includegraphics[width=0.33\textwidth]{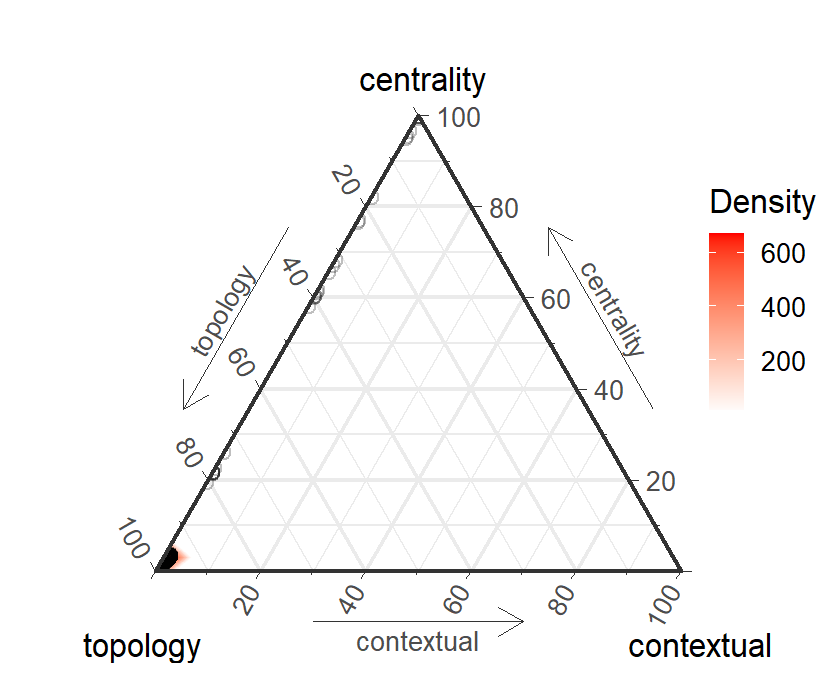}}
\subfloat[Dribbble]{\includegraphics[width=0.33\textwidth]{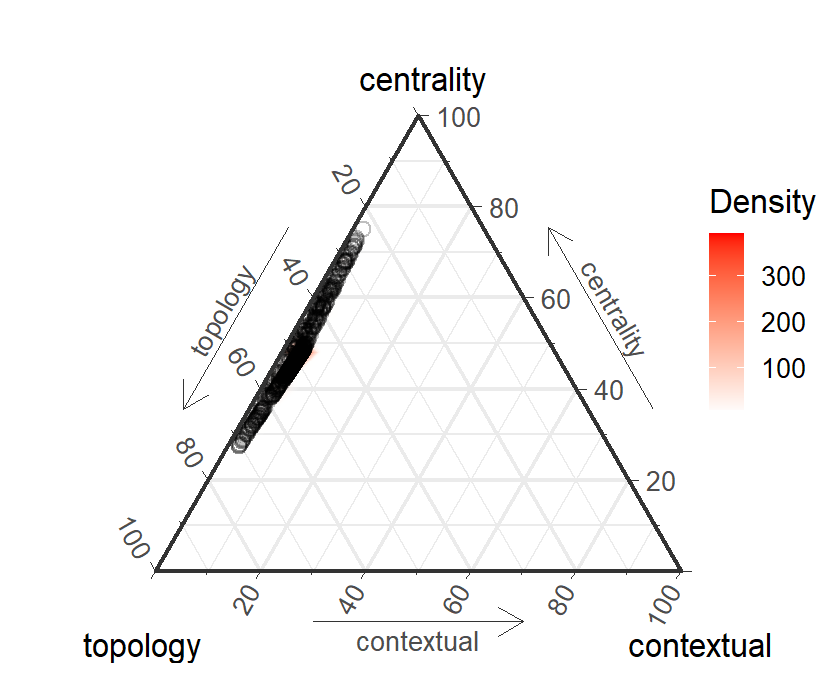}}
\subfloat[Kaggle]{\includegraphics[width=0.33\textwidth]{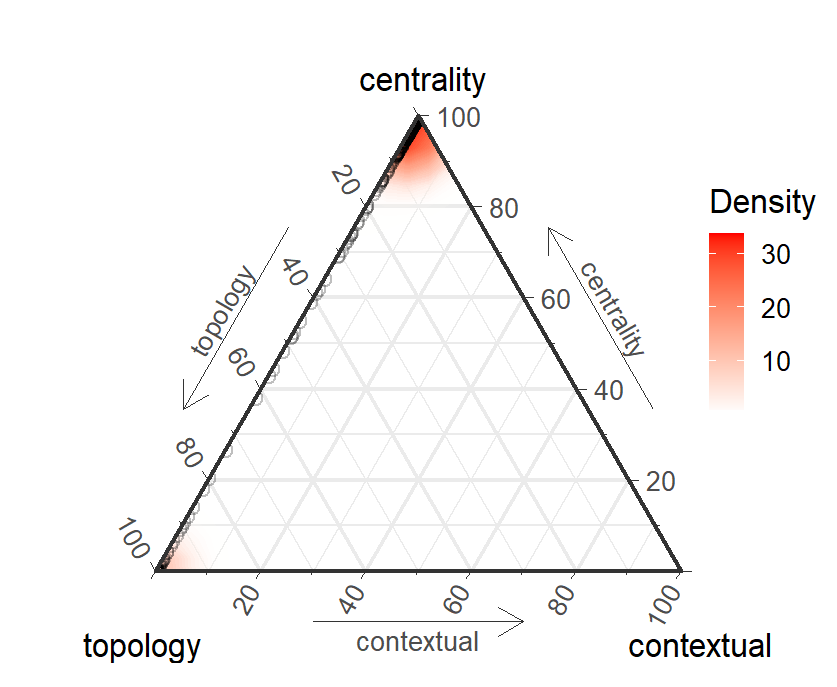}}

\vspace{-.30cm}

\subfloat[IMDb]{\includegraphics[width=0.33\textwidth]{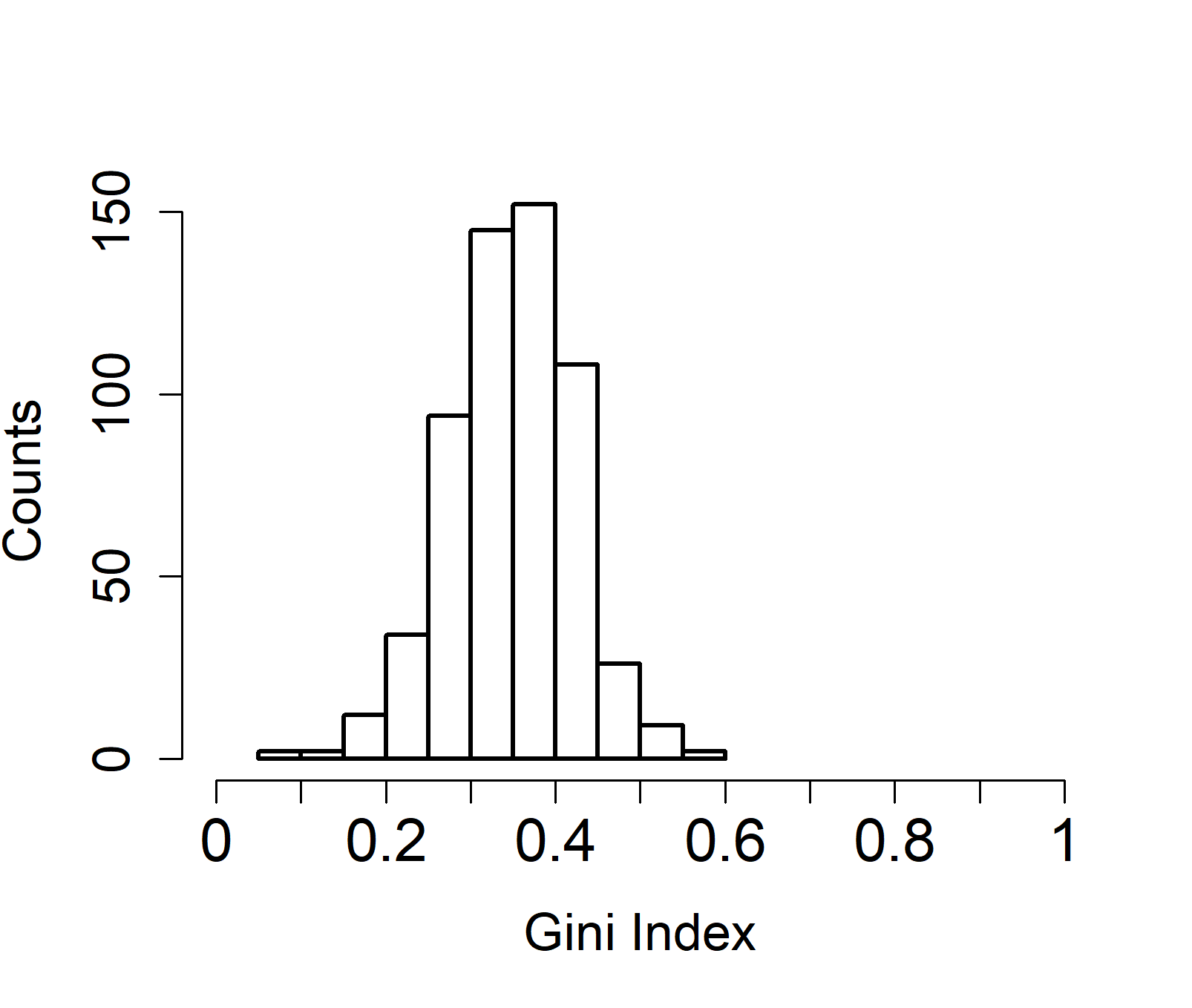}}
\subfloat[Dribbble]{\includegraphics[width=0.33\textwidth]{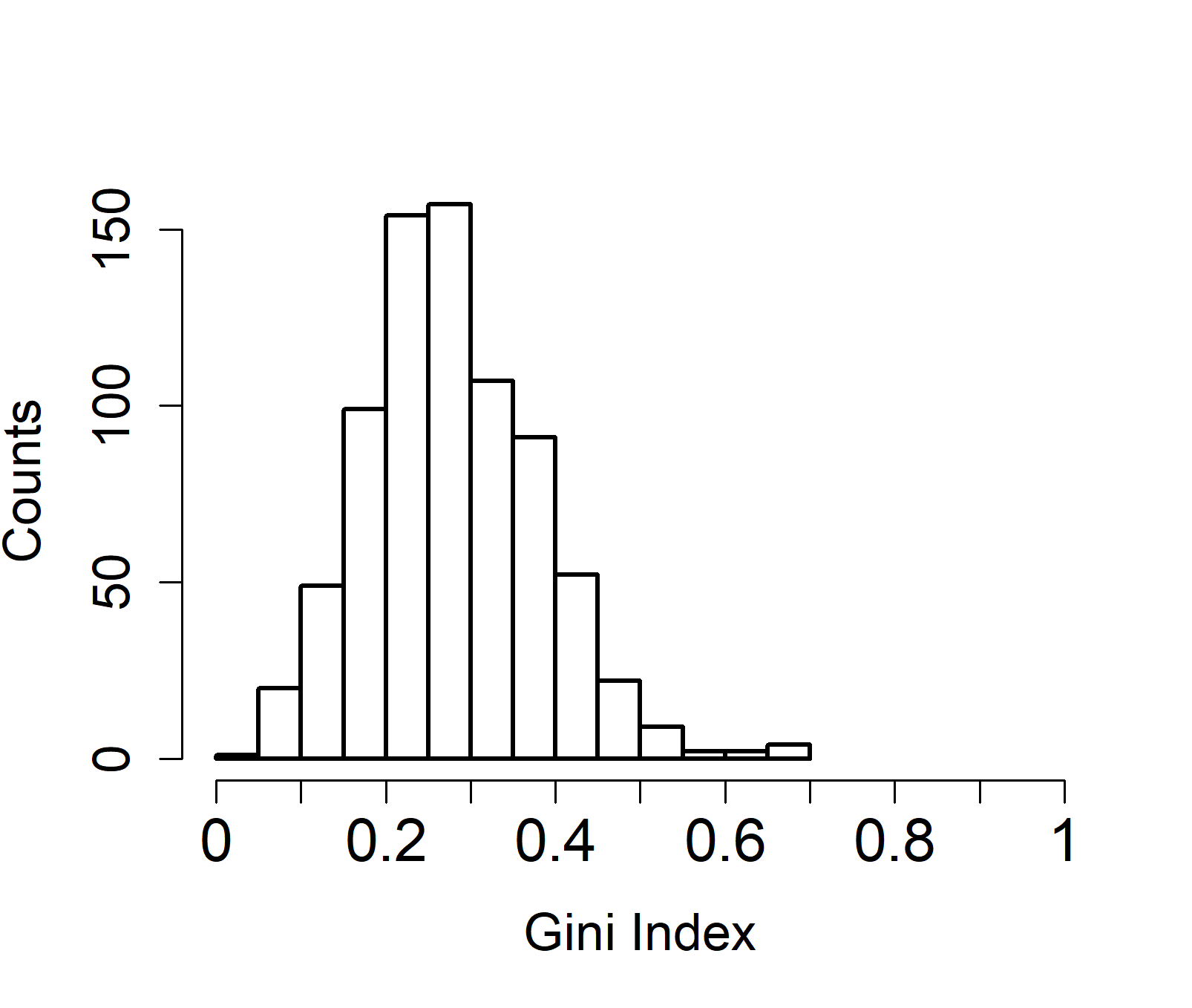}}
\subfloat[Kaggle]{\includegraphics[width=0.33\textwidth]{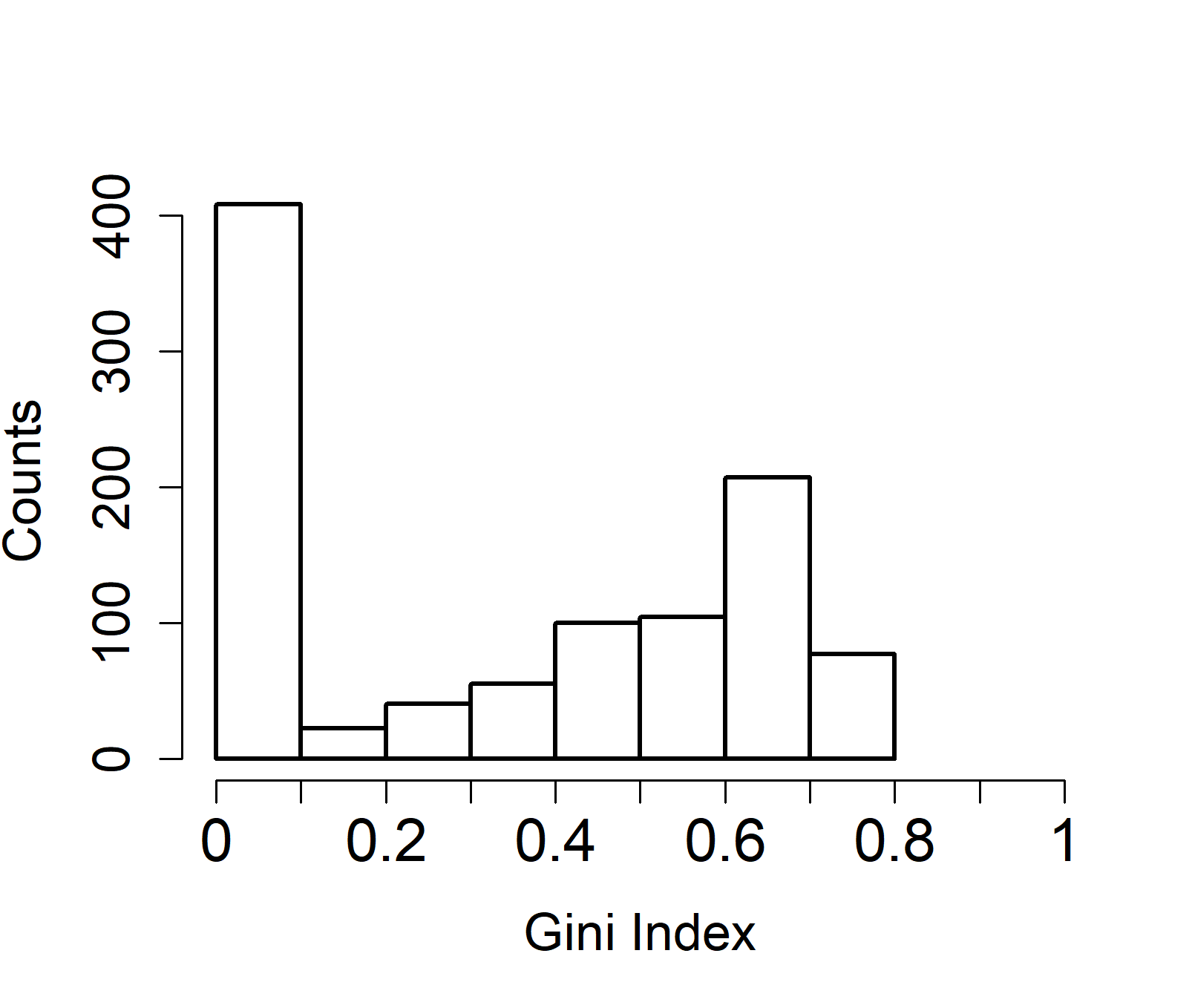}}

\caption[Optional caption for list of figures 5-8]{\textbf{The attention coefficients and Gini index on real-world datasets.} a-c) The attention coefficients explained in Formula~\ref{eq:attchannel} are visualized by means of ternary graphs; d-f) The distribution of the Gini index related to nodes' importance inside the teams. All the results refer to the configuration corresponding to the seed which returns the highest accuracy.}
\label{fig:att2}
\end{figure*}

\noindent
The attention mechanism of the 3-channels setting allows quantifying the contribution of each effect in determining the final outcome, fostering some degree of interpretability in an otherwise black-box model. Moreover, contributions of various channels can be visualized by resorting to ternary graphs. In our case, the ternary graph is populated with points, i.e., teams, whose location on the plot is given by attention weights of the three channels. Furthermore, by adding to the plot a 2D kernel density estimation, we try to address the problem of overplotting (i.e., many points over imposed on the same plotting area). In Figure~\ref{fig:attsy}, the effectiveness of the attention mechanism is shown on two synthetic datasets. The density of points across the plot recovers the rules used during the generating processes of the respective datasets: a) for L dataset, the model focuses its attention on contextual effects; b) in the case of LT dataset, a combination of the topological and contextual effects are considered.  

\noindent
Real-world scenario distributions of attention coefficients obtained by the aggregation mechanism are exposed in Figures~\ref{fig:att2}a-c. The findings show a diversified concentration of attention coefficients among different datasets and mostly no contribution from the contextual channel. In IMDb, topological effects seem to strongly drive the classification task. Let us remark that, being teams defined as cliques in IMDb, it is reasonable that nodes' attributes are key factors in determining team performance. In Dribbble, attention coefficients are evenly split between topology and centrality effects. This suggests that the connections a team has outside its workplace boost chances to reach the target audience, a critical factor given that Dribbble is a social media platform. In Kaggle, the centrality effect dominates the other ones, suggesting how co-working and shared ideas play an important role in determining a team's performance.

The attention mechanism defined at the node level in the first graph convolutional layer of the topology channel is able to pinpoint key nodes inside the team. This feature provides further insights that allow us to interpret the model's results. Firstly, we tested the effectiveness of this mechanism by designing a toy problem (for more details see Appendix~\ref{app:expressivity}). Secondly, we used node-level attention coefficients to spot the presence of "superstar" effects on real-world datasets. In other words, we tried to understand whether, in some teams, predictions were mostly driven by a unique node. Let us define the importance of each node as the sum of all the incoming attention coefficients according to the equation~\ref{eq:att2}, i.e.:

\begin{equation} \label{eq:importance}
    I_v = \sum_{j \in M} \alpha_{jv}
\end{equation}

\noindent
where $M$ denotes nodes connected by a directed edge pointing towards $v$ \footnote{We remark that directionality of message-passing procedure represent an hyper-parameter of the proposed model. In equation \ref{eq:importance} we assume that convolutions are done in a \textit{target-to-source} fashion. If instead convolutions are switched to \textit{source-to-target} the indexes in eq \ref{eq:importance} need to be switched}.

Now that we defined a metric to gauge a node's importance in a team, we are interested in understanding whether these contributions are evenly distributed inside the teams. We try to address this question by computing the Gini index of nodes' importances. The Gini Index is a measure of statistical dispersion whose application has grown beyond socioeconomic applications and reached various disciplines of science~\cite{gini1, gini2, gini3}. Crucially, the advantage of the Gini index is that it summarizes inequality in value distributions with a single scalar relatively simple to interpret. In more detail, the index takes values between 0 and 1, with 0 representing scenarios in which all of the nodes feature equal importance and 1 in scenarios in which there is only one very important node.

\begin{figure}[htp]
\centering
\subfloat[Attribute nodes]{\tikz\node[minimum height=\imageheight]{\includegraphics[height=5.7cm]{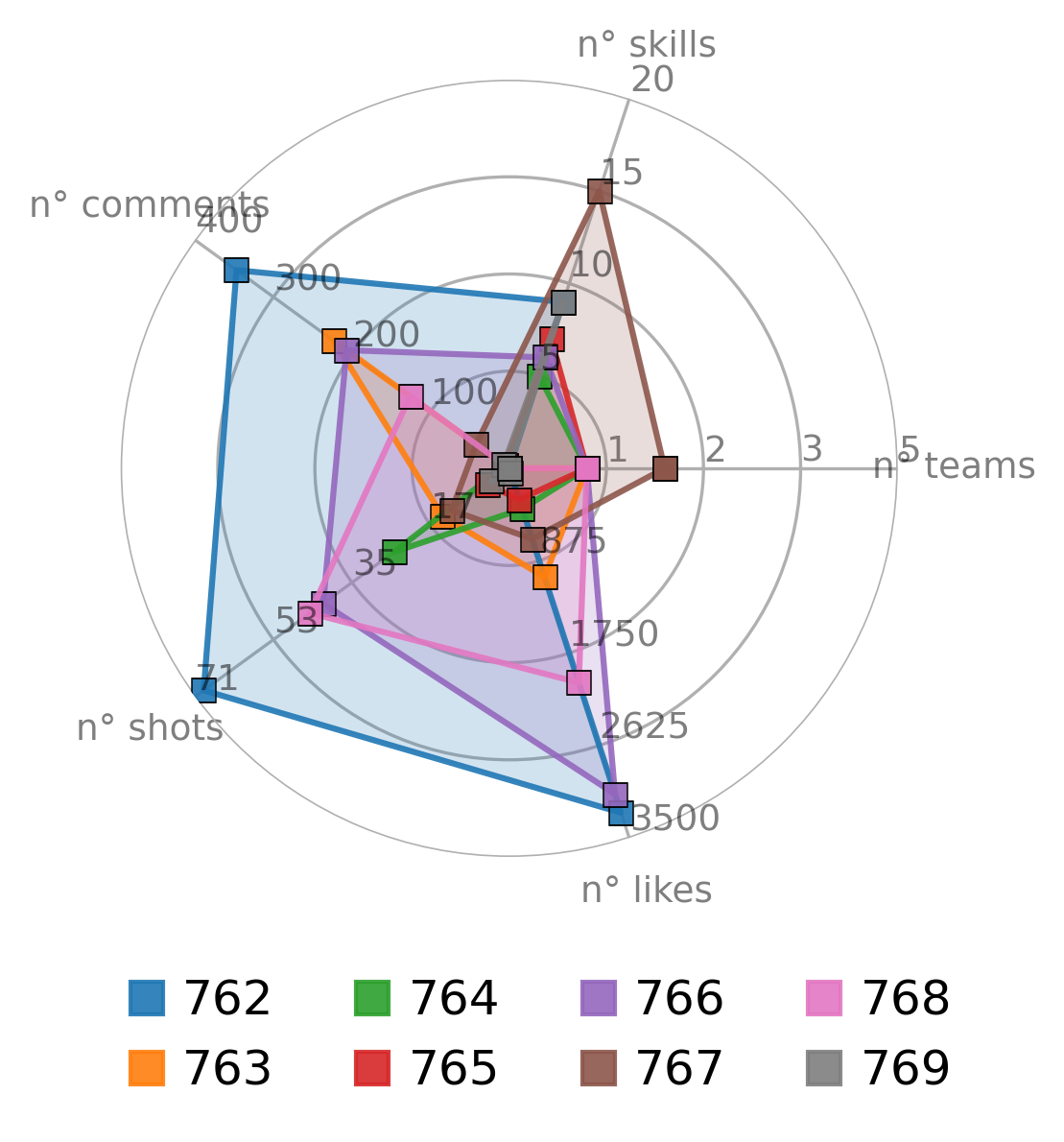}};}
\subfloat[Attention coefficients]{%
  \includegraphics[clip,width=0.4\columnwidth]{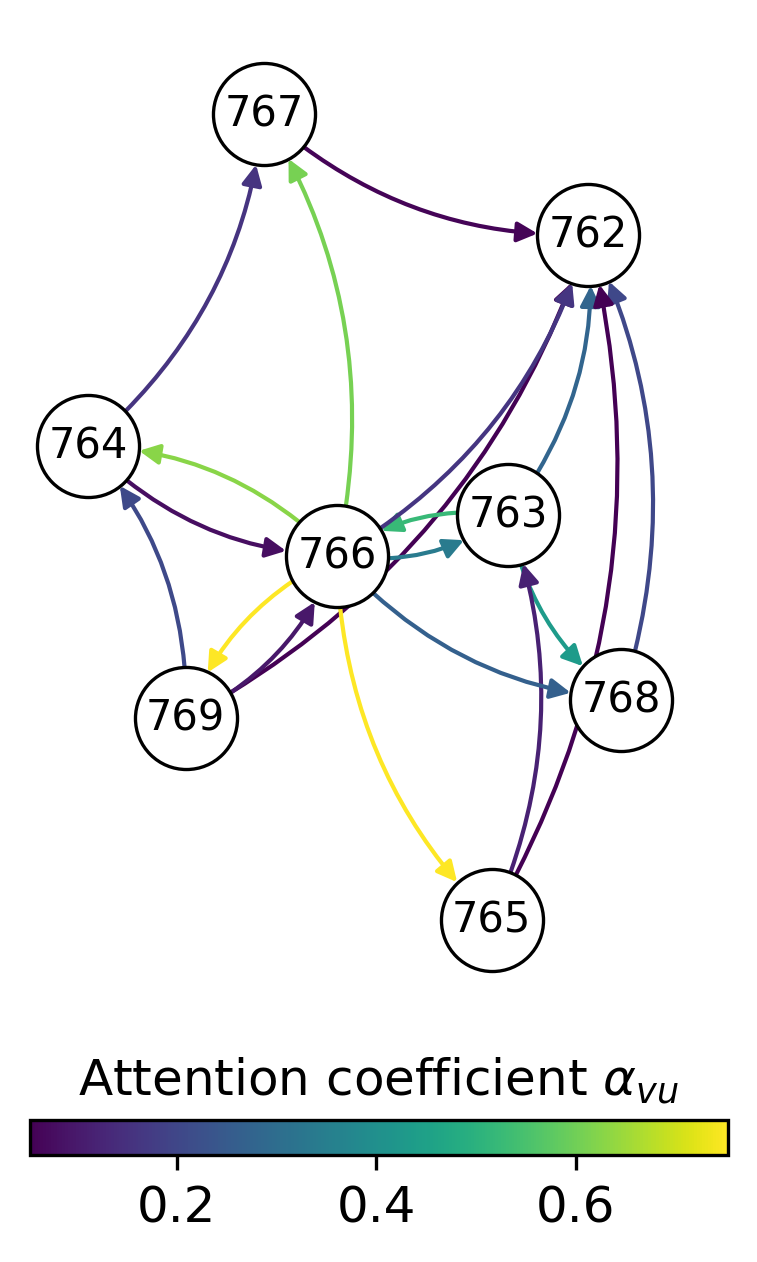}%
}
\caption{\textbf{Attention coefficients of the topology channel}. a) The values of the nodes' attributes of a team on Dribbble; b) The attention coefficients that the GATv2 layer returns at the topology level for a team on Dribbble.}
\label{fig:example}
\end{figure}

\noindent
Let us compute the Gini index for each team and show the distributions of such scores by means of histograms in  Figures~\ref{fig:att2}d-f. The distributions suggest how no dataset features teams with absolute inequalities (Gini values in the left neighborhood of 1). However, values around $0.5$ can highlight subgraphs where node importance is skewed towards a few nodes. In figure \ref{fig:example} we show an example of a team with a Gini index of  $0.52$. The team analyzed belongs to the Dribbble dataset and is high performing team. We see how attentions coefficients highlight user 766 as an important one. Crucially, this user seems to play an important role within the team despite not being the most "skilled" user: the non-one-sided connectivity pattern, together with their attributes, is what makes them important. Therefore, our model may be able to spot influential nodes by combining complex pattern encompassing both attributes and topology. On the Kaggle dataset we show how mainly homogenous contributions take place. 

\vspace{3mm}
\noindent
Concluding, let us remark on how the reliability of these findings increases with model performance (the higher the performance, the more reliable the attention coefficients). This consideration particularly holds with respect to the Kaggle dataset, on which performance results were not satisfying.

\section{Discussion}

\noindent
As already introduced in Section~\ref{sec:Ablation}, we observe how the proposed architecture is consistently capable of leveraging the attention mechanism and the preprocessing steps to focus on the meaningful effects driving the system. In more detail, even if we specify an architecture largely overparameterized where most of the parameters don't contribute to solving the final problem, the model is able to avoid over-fitting and generalize correctly to unseen data. It is important to stress how architectural parameters such as the number of convolutional layers in each channel were not object of the hyperparameter optimization procedure. On the one hand, choosing a tailored final model layout would probably enable us to push model performances further. On the other hand, we wanted to show that the proposed architecture was able to be robust even if overparameterized and with too many convolutional layers. The final results highlight therefore how the proposed architecture can be used out of the box on different datasets.

\noindent
It is important to highlight how in many contexts it is not easy to come up with a clear and well-defined notion of team performance. Furthermore, the performance might be influenced by many exogenous and external factors that might be hard to capture. Nevertheless, the experiments on synthetic data showed us that when the target quantity is directly a function of network properties, the model is able to correctly learn the underlying mechanisms. Therefore, definitions of performance in real scenarios closely related to network effects will likely be modeled more accurately by the proposed architecture.

\noindent
Lastly, let us remark on how edges in the input graph should encode interactions and social proximity between agents in a complex system. Considering that we try to model team performance by leveraging different kinds of network effects, we implicitly assume that edges convey predictive signals. These kinds of information are probably contained in high-resolution and privacy-sensitive databases of face-to-face interactions and private messaging logs which are not freely available for research purposes in most cases. When using social networks, we instead resort to encoding nodes' interactions by means of "follow" relations. This type of interaction may be seen as "socially weak" and not conveying strong enough signals to predict team performance. 

\section{Conclusion}

We present MENTOR, a framework for modeling team performances through neural graph representation learning techniques. MENTOR provides a tool to embed subgraphs belonging to a larger network leveraging concepts rooted in compilational models. In more detail, we propose different preprocessing steps and structural model features (i.e., 3-channels) to identify topological, centrality and contextual effects. Those effects are then aggregated by means of a soft-attention mechanism that provides both expressitivity and interpretability to the proposed model. In addition, the attention mechanism inside the topology channel provides further insights about nodes' importance inside the teams. We applied the model to ten datasets (7 synthetic and 3 real-world) representing teams dynamics in a graph structured system. Besides, the proposed synthetic datasets provide a new set of benchmarks tasks with respect to which future model expressivity could be tested. The proposed model largely outperforms all of the baselines on several synthetic tasks. Moreover, MENTOR outperforms the classical machine learning methods and the current neural baselines on real-word datasets, except for Kaggle dataset where poor performances occur for all the models. In addition, we stress how, in contrast with current neural baselines, MENTOR delivers straightforward interpretability by means of attention coefficients. This information can be useful in detecting the factors that affect performances in reference scenarios and the influence that the members of the teams exert when collaborating with each other.

\section*{Data and Code availability}

The data and code used to generate the results reported in this study will be made available at the time of publication.

\section*{Acknowledgments}
The research has been funded by the U. S. Army Research Laboratory and the U. S. Army Research Office under contract/grant number W911NF2010146. All authors thank Ciro Cattuto for useful discussions and suggestions.

\section*{Author contributions}
F.C. and P.F. collected the data, developed the computational pipeline, run the experiments, and wrote the first draft of the manuscript. N.P. and R.S. supervised the project. All authors conceived the research, interpreted the results, edited and approved the submitted version.


\begin{appendix}

\section{Further experiments on model capability}

\subsection{Further experiments on the direction of the convolution}
\label{app:direction_convolution}

\begin{figure}[ht!]
    \centering
    \includegraphics[width=0.7\linewidth]{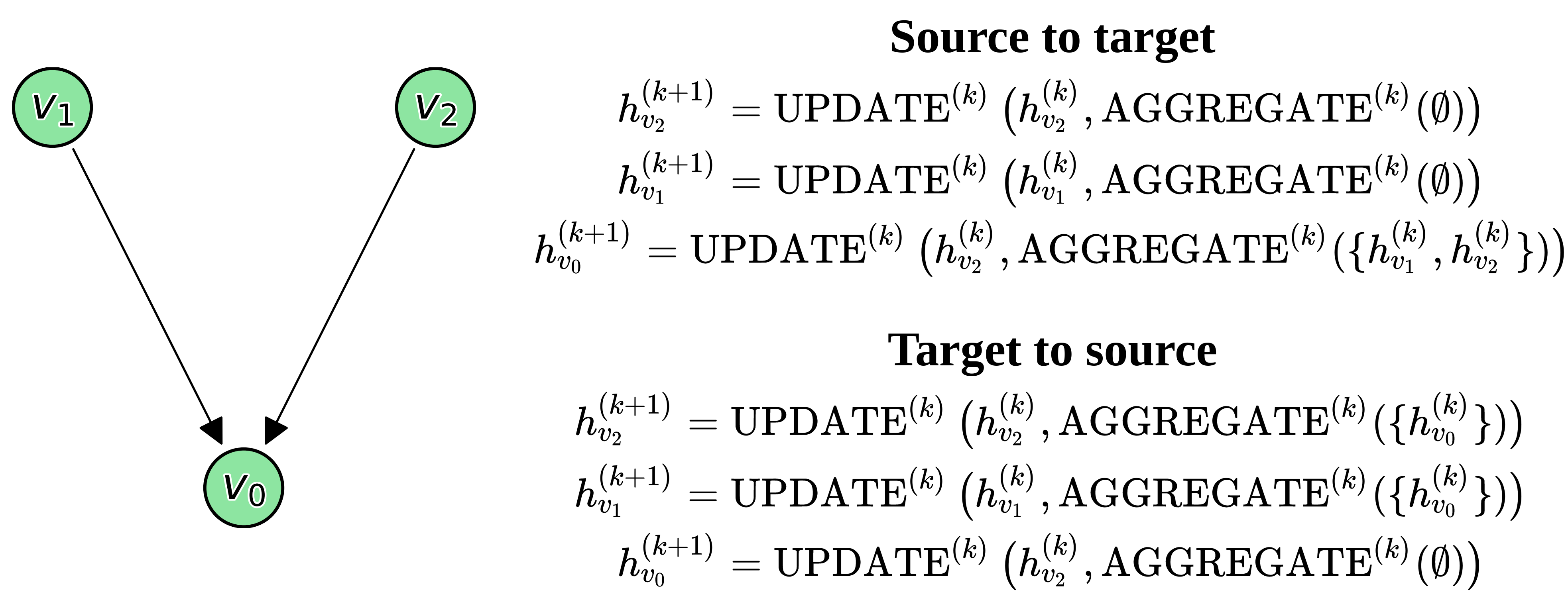}
    \caption{The choice regarding the direction of convolution influences the message passing procedure.}\label{fig:convdir}
\end{figure}

While constructing our model, we understood how seemingly harmless hyperparameters of the model play a fundamental role in determining the model expressivity. Chiefly, we remark how the \textbf{direction of the convolution} with respect to which nodes pool information during message passing procedure, plays a key role in capturing several important effects, as shown in Figure \ref{fig:convdir}. In more depth, we see that in centrality datasets the result of the MENTOR exposed in Table \ref{tab:results1} for CIn drops to $39.2 \%$ just by switching the direction of the convolution from \textit{source to target} to \textit{target to source} (the same behavior is seen with opposite direction in COut). Ignoring edge direction is probably what hinders SubGNN performance in tasks such as centrality.

\subsection{Further experiments on model expressivity and interpretability}
\label{app:expressivity}

As we remark in Section \ref{sec:topology}, the presence of the GAT layer inside the MENTOR architecture provides intra-teams explainability. In more details, the GAT attention mechanism allows getting coefficients that highilight and quantify the relative importance of internal connections. In order to provide an intuitive representation of this implementation, we created a directed graph where the teams' label is driven by the value of the \textit{attributes} of one member of the team. The graph is initially built by connecting $10\,000$ nodes through Erdos Renyi model. Next, we assigned each node in the network to a corresponding team within which the internal structure is defined using a clique. Finally, the nodes acquire attributes in order to discriminate the teams based on the attribute of a single node in the teams. We created the dataset in order to have 3 labels. For this reason, we have 4 types of nodes:

\begin{itemize}
    \item \textit{pro}: a particular node with \textit{high} skills;
    \item \textit{good}: a particular node with \textit{medium} skills;
    \item \textit{mediocre}: a particular node with \textit{low} skills; 
    \item \textit{noob}: a common node with basic skills. skills;
\end{itemize}

\begin{figure}[ht]
  \centering
  \subfloat[]{\includegraphics[width=0.5\textwidth]{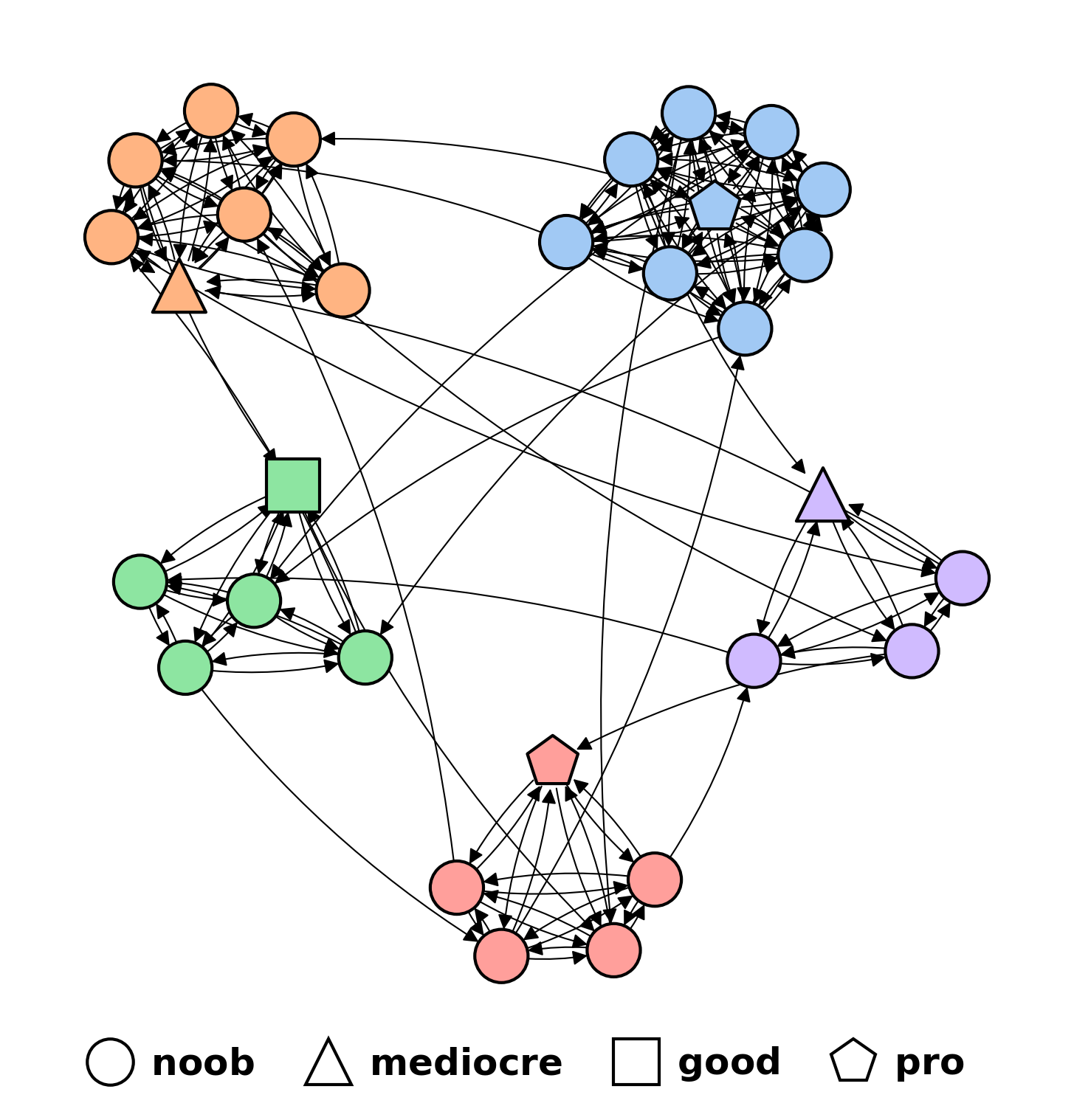}\label{fig:attr1}}
  \hfill
  \subfloat[]{\includegraphics[width=0.5\textwidth]{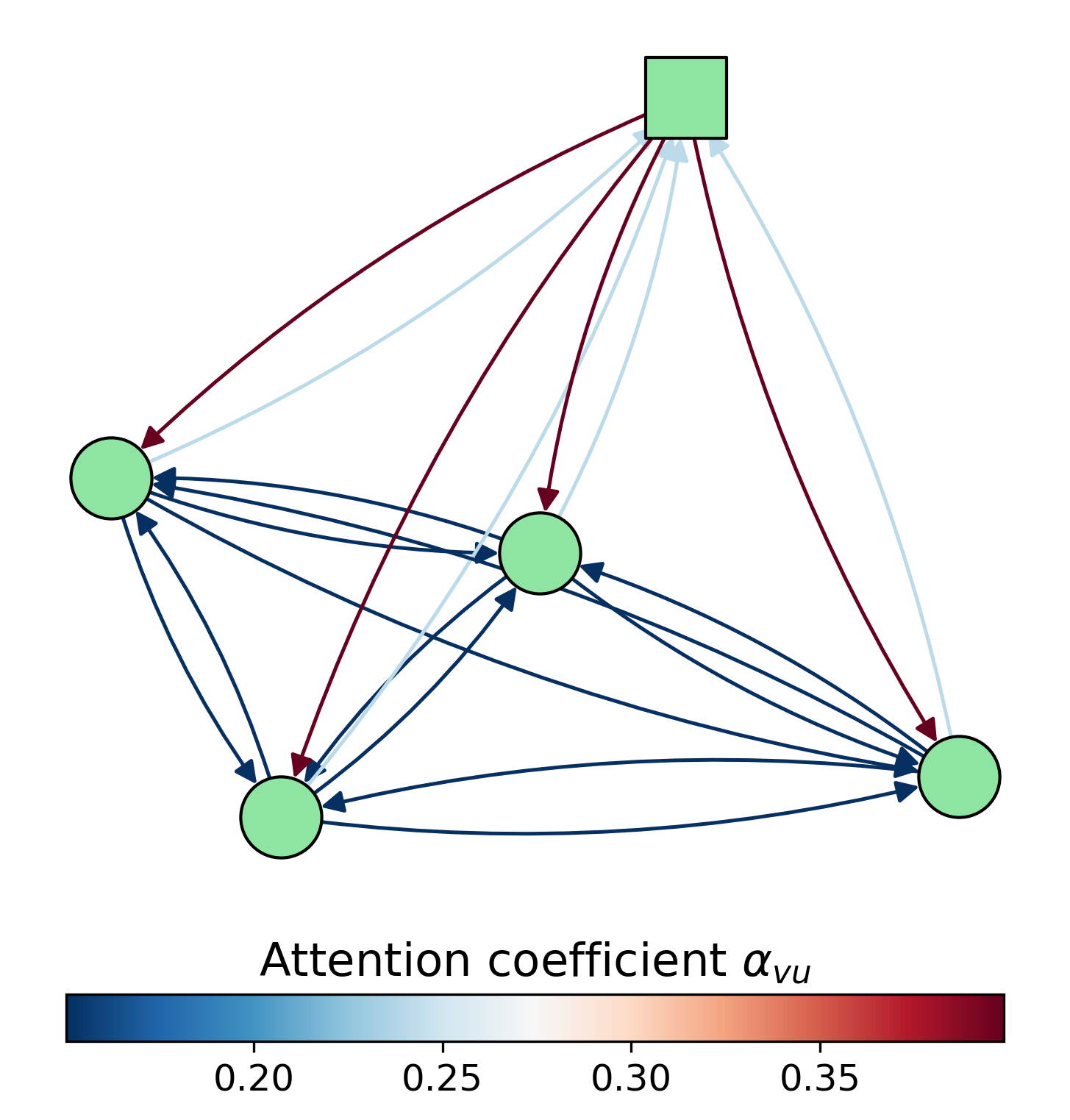}\label{fig:attr2}}
  \caption{(a) The resulting example graph for the explanation of GAT attention coefficients (the node colours identify the teams); (b) The attention coefficients resulting from training the MENTOR model on (a).}
\end{figure}

\noindent
An example of the output graph derived from the previous procedure is shown in Figure \ref{fig:attr1}.

\noindent
By looking at Figure \ref{fig:attr2}, we observe how after the training procedure of the model on the dataset proposed above, the GATv2 layer is able to correctly recover the discriminative node. In more depth, we see how edges related to the fundamental node exhibit attention coefficient quantitatively different from the other ones.

\section{Further Details on Data}
\label{app:data}

\subsection{Synthetic datasets}

\subsubsection{Frequently used algorithms}

Most of the synthetic datasets share some generative functions. In more details, they apply the following operations:

\begin{itemize}[noitemsep,topsep=0pt]
    \item Generation of the base graph with underlying noise structures (controlled by parameter $m$) given by directed Erdos-Renyi model (Algorithm \ref{alg:er});
    \item Team grouping procedure based on distribution sampling (Algorithm \ref{alg:team_maker});
    \item Generation of the subgraph internal structure by mean of predefined motif (Algorithm \ref{alg:MotifAdder}).
\end{itemize}

\noindent
For clarity of exposition, we report below the pseudo-codes of the above processes in order to refer to them later on.

\vspace{3mm}

\begin{algorithm}[H] 
\caption{DirectedErdosRenyi}
\label{alg:er}
\SetAlgoLined
\SetKwInOut{Input}{Input}
\SetKwInOut{Output}{Output}
\Input{Number of nodes $|\mathcal{V}|$; Number of edges $m$ created for each node;}
\Output{Graph $\mathcal{G}=(\mathcal{V},\mathcal{E})$;}
\BlankLine
$\mathcal{V} \leftarrow \{v_i:i=1,...,|\mathcal{V}|\}$

$\mathcal{E} \leftarrow \emptyset$

\For{$v_i \in \mathcal{V}$}{ 
    \For{$j = 1, ..., m$ }{
    \BlankLine
      $v_j \sim \operatorname{DiscreteUniform}(\mathcal{V} \setminus \{v_i\} )$
      
      $\mathcal{E} \leftarrow \mathcal{E} \cup \{(v_i,v_j)\}$
    \BlankLine
    }
}
\BlankLine

\Return{$\mathcal{G}=(\mathcal{V},\mathcal{E})$}

\end{algorithm}

\begin{algorithm}[H] 
\caption{TeamMaker}
\label{alg:team_maker}
\SetAlgoLined
\SetKwInOut{Input}{Input}
\SetKwInOut{Output}{Output}
\Input{Graph $\mathcal{G}=(\mathcal{V},\mathcal{E})$, Minimum team size $d_{min}$; Average incremental team size $\mu$;}
\Output{Partition $\mathcal{S}$ of subgraphs of variable size;}
\BlankLine
$\mathcal{V}_{available} \leftarrow \mathcal{V}$

$i \leftarrow 1$

\While{$\mathcal{V}_{available}$ $\operatorname{not}$ \  $\emptyset$}{ 
    
    \BlankLine
    $d_i \sim \operatorname{Poisson}(\mu)$ 
    
    $d_i \leftarrow d_{min} + d_i$
    
    \BlankLine
    
    $S_i \leftarrow (\mathcal{V}_{S_i} = \emptyset,\mathcal{E}_{S_i}=\emptyset)$
    
    \For{$j=1,...,d_i$}{
    
        $v_i \sim \operatorname{DiscreteUniform}(\mathcal{V}_{available})$
        
        $\mathcal{V}_{S_i} \leftarrow \mathcal{V}_{S_i} \cup \{v_i\}$
        
        $\mathcal{V}_{available} \leftarrow \mathcal{V}_{available} \setminus \{v_i\}$
    
    }
    
    $\mathcal{E}_{S_i} \leftarrow \{(v_k,v_l) \ | \ v_k,v_l \in \mathcal{V}_{S_i} \text{ and } (v_k,v_l) \in \mathcal{E} \}$
    
   $i \leftarrow i + 1$ 
}

\Return{$\mathcal{S} = \{S_1 = (\mathcal{V}_{S_1},\mathcal{E}_{S_1}),...,S_n = (\mathcal{V}_{S_n},\mathcal{E}_{S_n})\}$}, $n$ depend on the randomness of algorithm
\end{algorithm}

\vspace{3mm}
\noindent
The Algorithm \ref{alg:team_maker} should handle many corner case scenario that we do not report in the pseudo-code for clarity of explanation. Refer to the code in the Github repository for the actual implementation.

\vspace{3mm}
\noindent
Let us now remark how a generic motif $\mathcal{M}_l$ is defined as a set of specific edges:

$$\mathcal{M}_l=\{e_1,...,e_h\}$$

\begin{figure*}[ht]
\centering
  \includegraphics[scale=0.7]{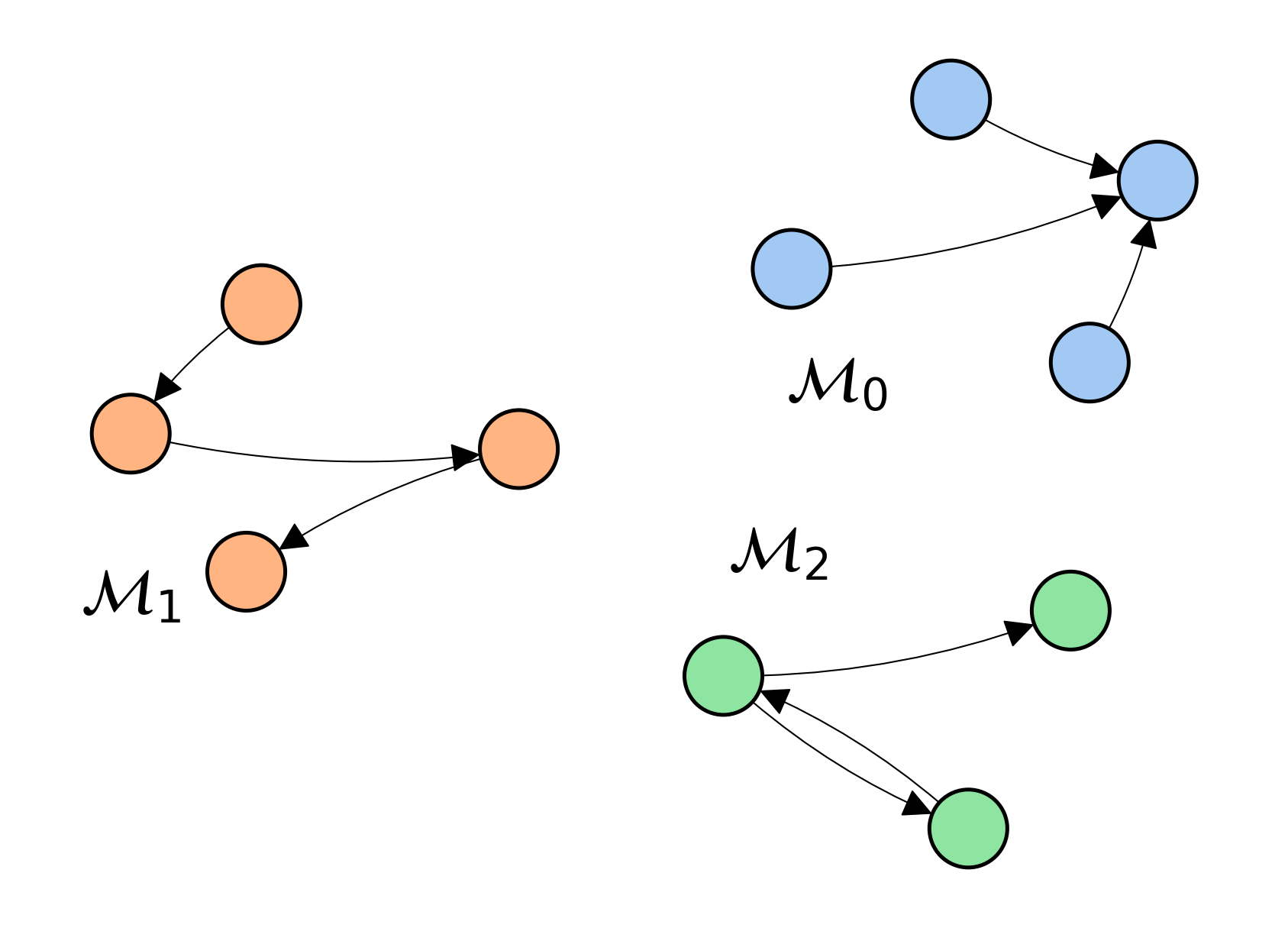}
  \caption{\textbf{Motifs}. Three types of connectivity patterns, a.k.a \textit{motifs}.}
  \label{fig:Motifs}
\end{figure*}

\noindent
Let us denote with $\mathcal{P}(A)$ the powerset of a generic set $A$. We define functions $B_{\mathcal{M}_l}: \mathcal{P}(\mathcal{V})  \rightarrow \mathcal{P}(\mathcal{V} \times \mathcal{V})$ such that, given $\{v_1,...,v_k\} \subset V_{S_i}$, $B_{\mathcal{M}_l}$ returns the connectivity pattern specified in ${\mathcal{M}_l}$ using nodes in $\{v_1,...,v_k\}$. Let us remark that $k$ is the number of nodes involved in the motif and may vary from motif to motif ($|\mathcal{M}_l|=k$). For example, let us observe ${\mathcal{M}_0}$ in Figure \ref{fig:Motifs}. By sampling nodes without replacement $\{v_1,v_2,v_3,v_4\} \subset \mathcal{V}_{S_i} \subset \mathcal{V}$, we have:

$$B_{\mathcal{M}_0}(\{v_1,v_2,v_3,v_4\}) = \{(v_2,v_1),(v_3,v_1),(v_4,v_1)\}$$

\begin{algorithm}[H] 
\caption{MotifAdder}
\label{alg:MotifAdder}
\SetAlgoLined
\SetKwInOut{Input}{Input}
\SetKwInOut{Output}{Output}
\Input{Graph $\mathcal{G}=(\mathcal{V},\mathcal{E})$; A team $S_i = (\mathcal{V}_{S_i},\mathcal{E}_{S_i}) \in \mathcal{S}$; Ratio of nodes $r$ involved into the repeatedly addition of motifs; Type of motif to add $\mathcal{M}_l$;}
\Output{Graph $\mathcal{G}=(\mathcal{V},\mathcal{E})$ with motif structures added to the team $S_i$;}
\BlankLine

\BlankLine
\For{$j = 1, ..., \lfloor |\mathcal{V}_{S_i}|* r \rfloor$ }{
\BlankLine
  
  $v_1,...,v_k \sim \operatorname{DiscreteUniform}(\mathcal{V}_{S_i})$ without replacement \texttt{   // where $k = |\mathcal{M}_l|$ }
  
  $\mathcal{E} \leftarrow \mathcal{E} \cup B_{\mathcal{M}_l}(\{v_1,...,v_k\})$ 
  
  $\mathcal{E}_{S_i} \leftarrow \mathcal{E}_{S_i} \cup B_{\mathcal{M}_l}(\{v_1,...,v_k\})$
\BlankLine
}
\BlankLine

\Return{$\mathcal{G}=(\mathcal{V},\mathcal{E})$; $S_i = (\mathcal{V}_{S_i},\mathcal{E}_{S_i})$;}

\end{algorithm}

\subsubsection{Centrality} \label{app:centrality}

\begin{figure}[ht]
  \centering
  \subfloat[Graph $\mathcal{G}=(\mathcal{V},\mathcal{E})$.]{\includegraphics[width=0.333\textwidth]{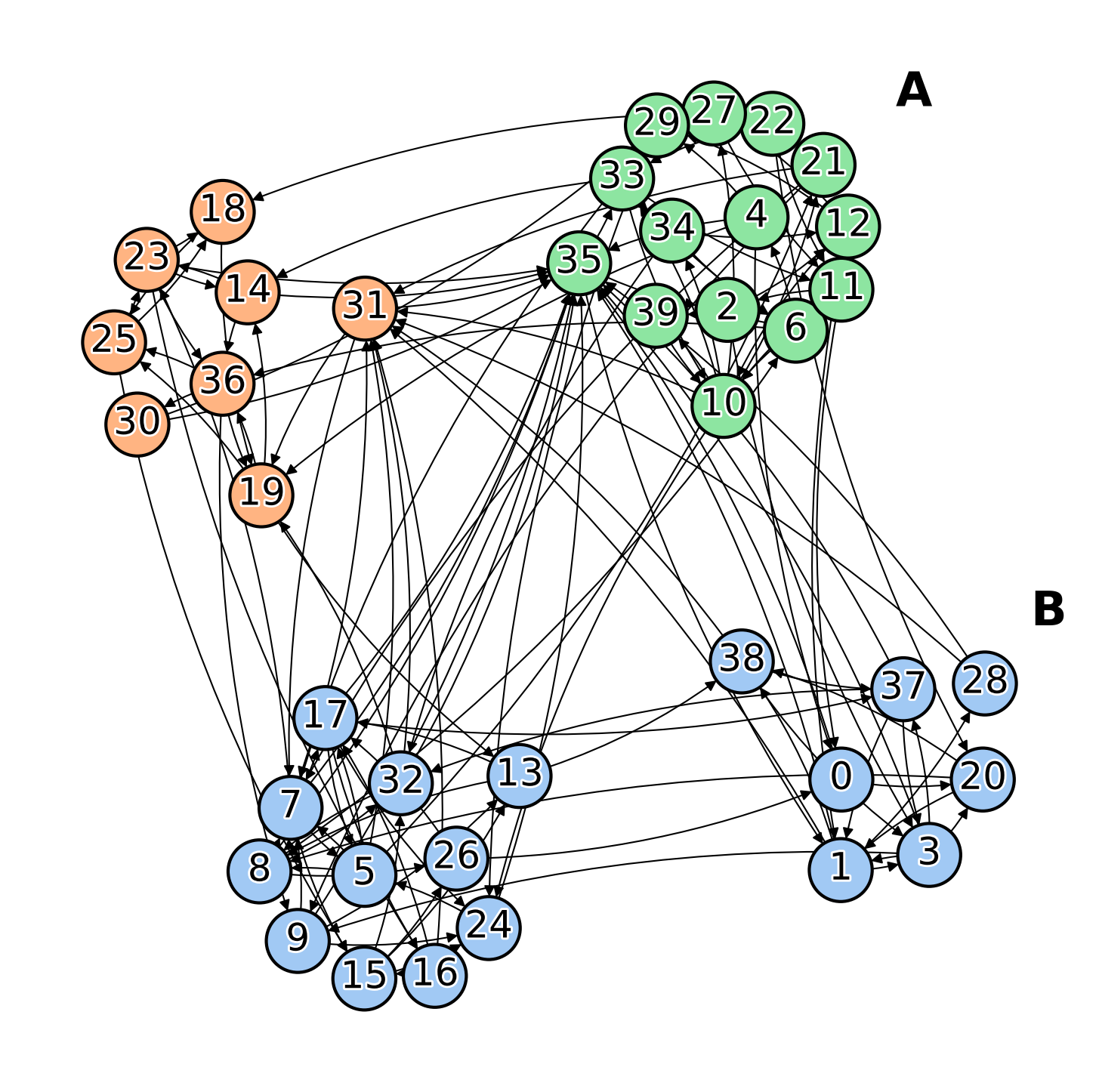}\label{fig:c1}}
  \hfill
  \subfloat[Team A discrimination.]{\includegraphics[width=0.333\textwidth]{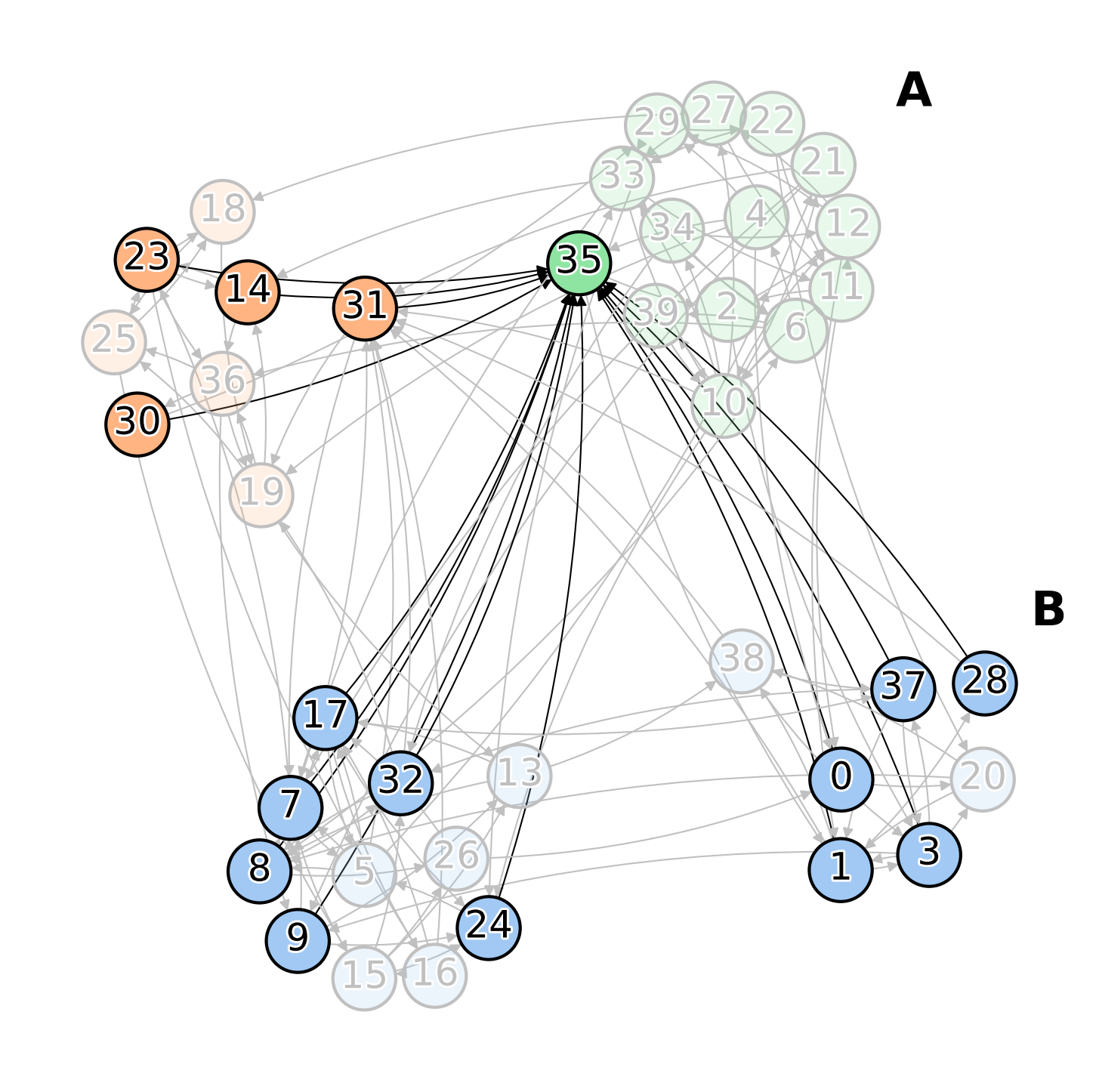}\label{fig:c2}}
  \hfill
  \subfloat[Team B discrimination.]{\includegraphics[width=0.333\textwidth]{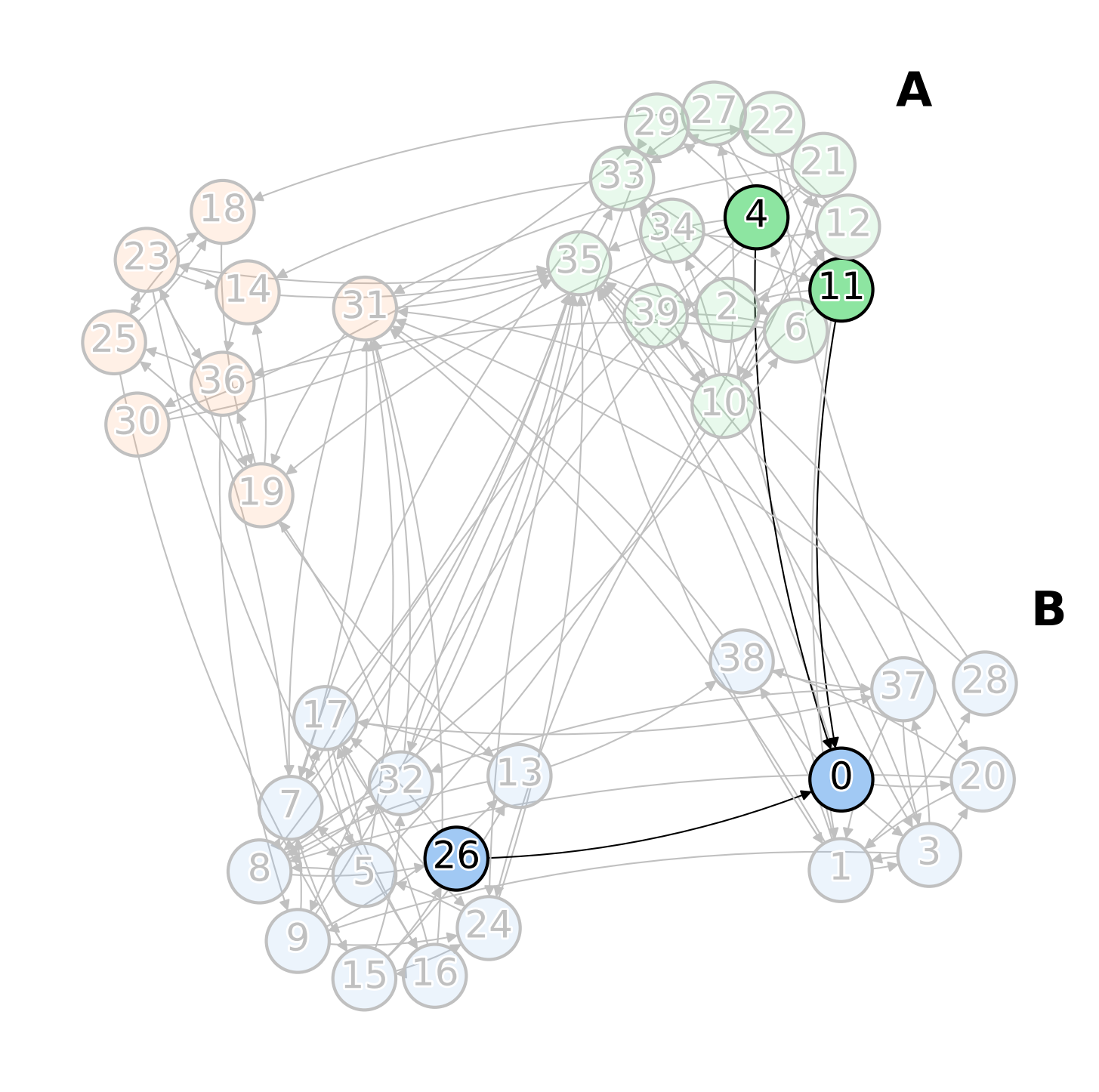}\label{fig:c3}}
  \caption{\textbf{Centrality}. The node colours identify the labels of the teams, $C = \{0, 1, 2\}$. (a) The resulting example graph from Algorithm \ref{alg:CIn}. (b) The discriminative node 35 of the team A has an high in-degree from the outside: the team A have label 2. (c) The node 0 of the team B has a small in-degree from the outside: the team have label 0.}
  \label{fig:Centrality}
\end{figure}

We designed centrality datasets aiming to assess model capabilities in predicting teams' labels based on the overall  \textit{in-degree} and \textit{out-degree} of the subgraph representing them. In more detail, we verified whether the model was able to discover that the final team's label was determined by the connectivity structure of a randomly selected member of the team\footnote{We release code that can be used to assess scenarios in which links are assigned with different strategies by leveraging a sampling mechanism based on the Dirichlet distribution.}. Algorithm \ref{alg:CIn} exposes the operational phases through which the graph is obtained for the "in-degree" scenario (see Figure \ref{fig:Centrality} for a simplified representation of the resulting graph). We remark how the "out-degree" case can be derived by changing the order direction in Lines \ref{line:c1} and \ref{line:c2} of the Algorithm \ref{alg:CIn}.

\noindent
For both "in-degree" and "out-degree" cases, we use the following set of parameters: $|\mathcal{V}| = 10\,000$, $m = 1$, $d_{min} = 5$, $\mu = 5$, $r = 0.8$, $C = 3$ and $\delta = 20$.

\begin{algorithm}[H] 
\caption{Centrality}
\label{alg:CIn}
\SetAlgoLined
\SetKwInOut{Input}{Input}
\SetKwInOut{Output}{Output}
\Input{Number of nodes $|\mathcal{V}|$; Number of edges $m$ created for each node; Minimum team size $d_{min}$; Average incremental team size $\mu$; Ratio of nodes $r$ involved into the repeatedly addition of motifs; An unique motif $\mathcal{M}_1$; Number of classes $C$; Separation value $\delta$ between centers of the mass of probability distributions generating the classes;}
\Output{Graph $\mathcal{G}=(\mathcal{V},\mathcal{E})$; Set of subgraphs $\mathcal{S}=\{S_1,S_2,...,S_n\}$; Set of labels for each subgraph, $\mathcal{Y}=\{y_{S_i} \ | \ S_i \in \mathcal{S}$ \};}
\BlankLine

$\mathcal{G} \leftarrow \operatorname{DirectedErdosRenyi}(|\mathcal{V}|,m)$

$\mathcal{S} \leftarrow \operatorname{TeamMaker}(\mathcal{G}, d_{min}, \mu)$

$\mathcal{Y} \leftarrow \emptyset$

$centers \leftarrow [\delta,2\delta,...,+C\delta]$

\BlankLine
\For{$S_i \in \mathcal{S}$}{ 
      
      $\mathcal{G},S_i \leftarrow \operatorname{MotifAdder}(\mathcal{G},S_i, r, \mathcal{M}_1)$ \texttt{   // updates $S_i \in \mathcal{S}$ }
      
}
\BlankLine

\For{$S_i \in \mathcal{S}$}{ 
      
      \BlankLine
      
      $j \sim \operatorname{DiscreteUniform}(\{0,...,C\})$
      
      \BlankLine
      \If{$j  \operatorname{not} C$}{
      
          $n_{links} \sim \operatorname{Normal}(centers[j],1)$
   
          $v \sim \operatorname{DiscreteUniform}(\mathcal{V}_{S_i})$
          
          $u_1,...,u_{n_{links}} \sim \operatorname{DiscreteUniform}(\mathcal{V} \setminus \mathcal{V}_{S_i})$ without replacement 
          
          $\mathcal{E} \leftarrow \mathcal{E} \cup \{(v,u_1),...,(v,u_{n_{links}})\}$  \label{line:c1}
      
          $\mathcal{E}_{S_i} \leftarrow \mathcal{E}_{S_i} \cup \{(v,u_1),...,(v,u_{n_{links}})\}$ \label{lst:line:blah2} \label{line:c2}
      
      }
      
      \BlankLine
      
      \Else{
      
          $n_{links} \sim \operatorname{DiscretePowerLaw}_{\alpha=3}(centers[j])$
   
          $v \sim \operatorname{DiscreteUniform}(\mathcal{V}_{S_i})$
          
          $u_1,...,u_{n_{links}} \sim \operatorname{DiscreteUniform}(\mathcal{V} \setminus \mathcal{V}_{S_i})$ without replacement 
          
          $\mathcal{E} \leftarrow \mathcal{E} \cup \{(v,u_1),...,(v,u_{n_{links}})\}$ 
      
          $\mathcal{E}_{S_i} \leftarrow \mathcal{E}_{S_i} \cup \{(v,u_1),...,(v,u_{n_{links}})\}$}{}
      
    $\mathcal{Y} \leftarrow \mathcal{Y} \cup \{j\}$
 
}
\BlankLine

\Return{$\mathcal{G}=(\mathcal{V},\mathcal{E})$; $\mathcal{S}=\{S_1,S_2,...,S_n\}$; $\mathcal{Y}=\{y_{S_i} \ | \ S_i \in \mathcal{S}\};$}

\end{algorithm}

\subsubsection{Topology} \label{app:Topology}

Topology datasets were designed to gauge the proposed framework's ability to discriminate teams by leveraging regular patterns in their connectivity internal topological structure. We provide three different implementations of topological datasets in order to evaluate models' performance on different nuances and specifications of topological structures. The tested synthetic datasets can be generated following the logical steps defined in Algorithm \ref{alg:T1}. More precisely, we outline below procedural details to obtain dataset T1. As regard dataset T2, the topological process is expanded with the preferential attachment procedure of Algorithm \ref{alg:CIn}, without involving the corresponding steps of class discrimination. The dataset T3 differs from T1 by modifying the Algorithm \ref{alg:team_maker} to allowing the generation of overlapping teams and nodes without any teams. Another relevant difference between these datasets is dictated by the noise effect regularized through parameter $m$. In fact, $m$ is set to 5 in T1 and T3 while is set to 1 in T2.

\noindent
For all the topological datasets, we use the following set of parameters: $|\mathcal{V}| = 10\,000$, $d_{min} = 5$, $\mu = 5$ and $r = 0.8$.

\vspace{3mm}
\begin{algorithm}[H] 
\caption{T1}
\label{alg:T1}
\SetAlgoLined
\SetKwInOut{Input}{Input}
\SetKwInOut{Output}{Output}
\Input{Number of nodes $|\mathcal{V}|$; Number of edges $m$ created for each node; Minimum team size $d_{min}$; Average incremental team size $\mu$; Ratio of nodes $r$ involved into the repeatedly addition of motifs; An array of three motifs $M = [\mathcal{M}_0,\mathcal{M}_1,\mathcal{M}_2 ]$;}
\Output{Graph $\mathcal{G}=(\mathcal{V},\mathcal{E})$; Set of subgraphs $\mathcal{S}=\{S_1,S_2,...,S_n\}$; Set of labels for each subgraph, $\mathcal{Y}=\{y_{S_i} \ | \ S_i \in \mathcal{S}$ \};}
\BlankLine

$\mathcal{G} \leftarrow \operatorname{DirectedErdosRenyi}(|\mathcal{V}|,m)$

$\mathcal{S} \leftarrow \operatorname{TeamMaker}(\mathcal{G}, d_{min}, \mu)$

\BlankLine

$\mathcal{Y} \leftarrow \emptyset$

\For{$S_i \in \mathcal{S}$}{ 
      
      $j \sim \operatorname{DiscreteUniform}(\{0,1,2\})$
      
      $\mathcal{M} \leftarrow  M[j]$
      
      $\mathcal{G},S_i \leftarrow \operatorname{MotifAdder}(\mathcal{G},S_i, r, \mathcal{M})$ \texttt{   // updates $S_i \in \mathcal{S}$ }
      
     $\mathcal{Y} \leftarrow \mathcal{Y} \cup \{j\}$
}
\BlankLine

\Return{$\mathcal{G}=(\mathcal{V},\mathcal{E})$; $\mathcal{S}=\{S_1,S_2,...,S_n\}$; $\mathcal{Y}=\{y_{S_i} \ | \ S_i \in \mathcal{S}\};$}

\end{algorithm}

\subsubsection{Contextual}

\begin{figure*}[ht]
\centering
  \includegraphics[scale=0.7]{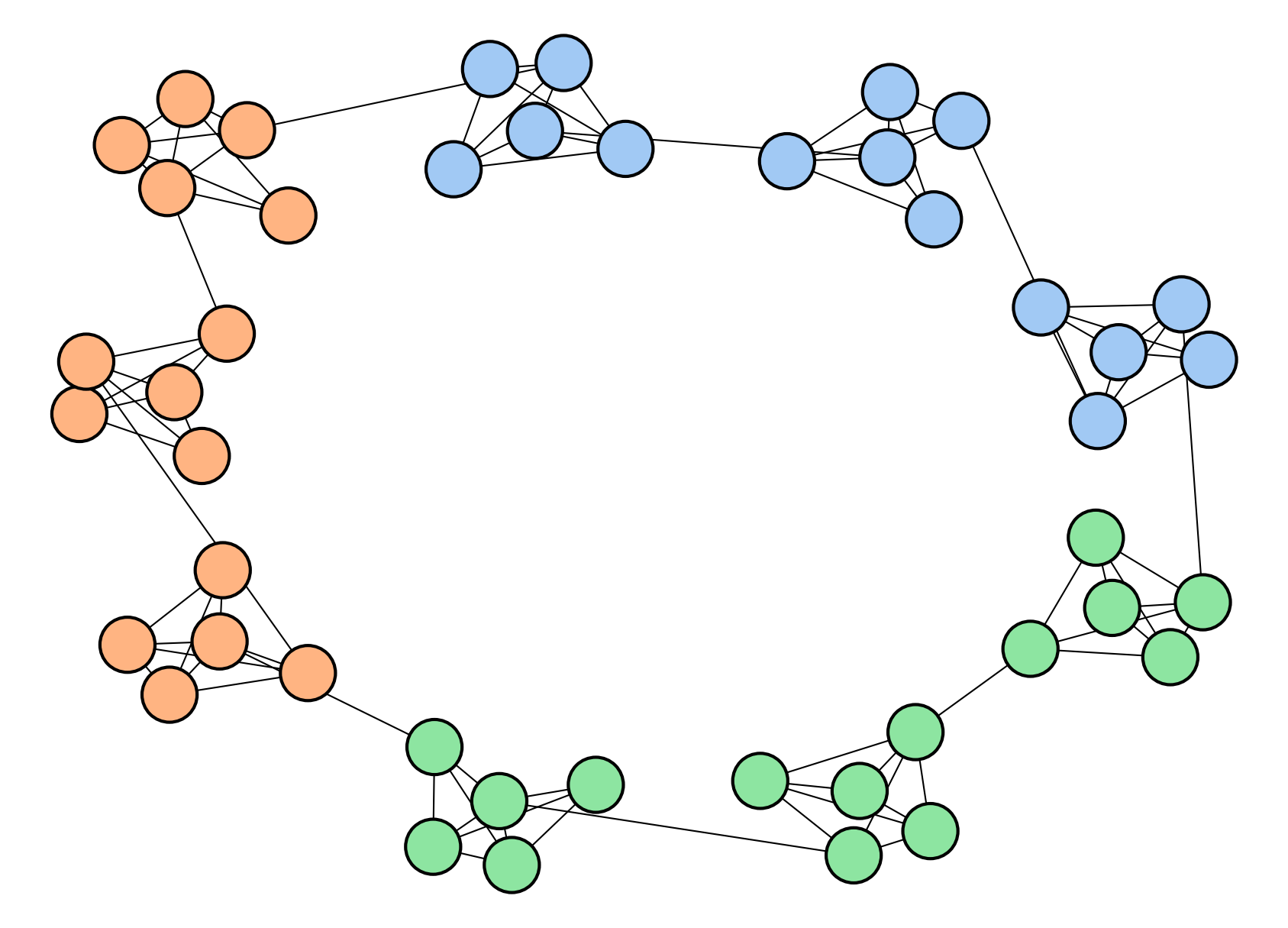}
  \caption{\textbf{Contextual.} The Caveman graph is divided into three parts that identify the label of the teams. The node colours identify the labels of the teams, $C = \{0, 1, 2\}$.}
  \label{fig:Position}
\end{figure*}

This synthetic dataset was designed in order to verify model expressivity with respect to contextual effects. In more depth, the aim is to gauge the model's ability to leverage regularities in graph structures that may be several hops apart in term of degrees of separations. The contextual dataset features an undirected graph composed by $n = 200$ cliques (teams) of size $k = 10$. Crucially, we leverage the idea of Caveman Graph \cite{caveman} in which a ring-shaped graph is obtained by bridging fully connected cliques through the rewiring of a single edge in each clique to a node in an adjacent clique. The idea is to split the ring-shaped network into three part and label each team based on their position in the graph (see Figure \ref{fig:Position}).

\subsubsection{Contextual \& Topology} \label{app:position_topology}

\begin{figure*}[ht]
\centering
  \includegraphics[width=\textwidth]{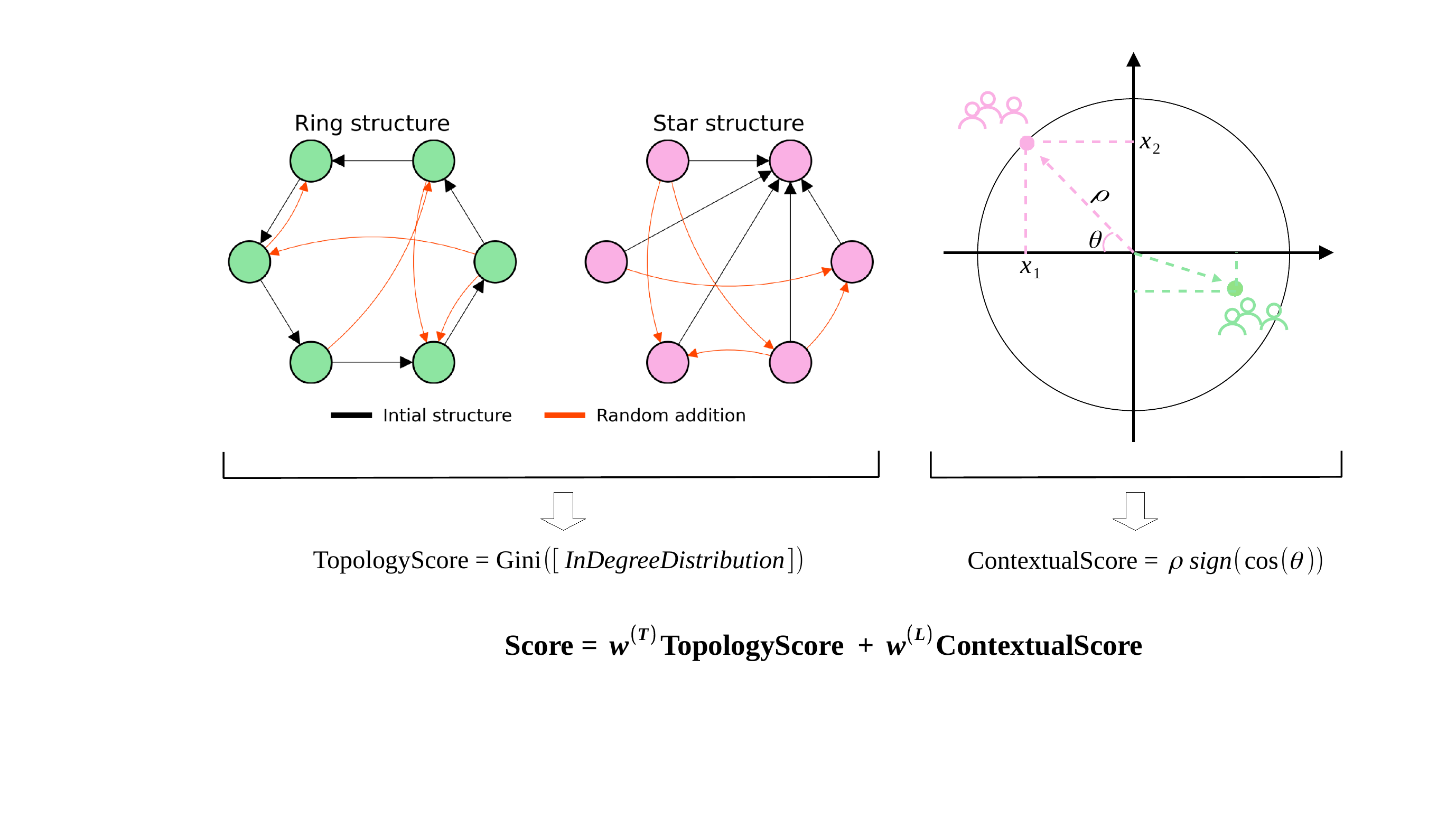}
  \caption{\textbf{Contextual\&Topology}. On the left, the two patterns of connectivity for the internal structure of the teams; on the right, the generation procedure of the cartesian coordinates for each team.}
  \label{fig:PositionTopology}
\end{figure*}

This synthetic dataset was designed to asses the model ability to function in scenarios where the final outcome is determined by the interaction of two effects: \textit{contextual} and \textit{topological}. We built $1000$ teams with variable sizes according to two different blueprints: ring-shaped and star-shaped graphs. The internal structures of both prototypes are then perturbed by randomly adding some edges. During subgraph generation, teams are also endowed with a pair of cartesian coordinates (x, y) sampled from a uniform distribution defined on a circle of radius 1 in $\mathbb{R}^2$. On the one hand the topological score is obtained by computing the gini index of the in-degree distribution of nodes inside the subgraph. In more detail, star-shaped topology will feature a high gini index whereas ring-shaped structures will feature low score. On the other hand, we first pass to the polar coordinate system and obtain the contextual score as $\rho\operatorname{sign}(\operatorname{cos}(\theta))$. The scores are then normalized through a standard scaling procedure and then merged through a equally weighted average\footnote{The code allows to specify different weights $w^{(T)}$ and $w^{(L)}$.}. These steps are shown in Figure \ref{fig:PositionTopology}. Concluding, in order to preserve the contextual effect generated through the sampling procedure exposed above, the inter-team connectivity structure is obtained by connecting subgraphs that are nearby in the cartesian space. Procedurally, we compute for each team its 6-nearest neighbors and then we randomly connect nodes belonging to the involved teams.

\subsection{Real datasets}
\label{app:Real datasets}

\subsubsection{IMDb}

We downloaded the raw dataset from Kaggle website: \url{https://www.kaggle.com/stefanoleone992/imdb-extensive-dataset}. In particular, we only leveraged information contained in several tables: \textit{IMDb movies.csv}, \textit{IMDb names.csv} and \textit{IMDb title principals.csv}. 

\noindent
In order to frame the problem of team performance prediction as a graph representation learning task, we restructure the available data according to several simplifying assumptions:

\begin{itemize}
    \item To obtain meaningful internal team connectivity structure, we retained only movies with at least four cast components;
    \item We only considered movies produced in USA with worldwide gross income evaluated in dollars;
    \item We only considered movies after 2018.
\end{itemize}

\noindent
The connectivity of nodes in the graph is defined by "co-working" relations between cast components. In more detail, we will consider as "team" the cast of a movie and define the internal connectivity structure as a clique. The resulting undirected graph is populated by $4\,802$ nodes and $25\,632$ edges. The nodes are partitioned into $586$ variable size and possibly overlapping teams. We considered as nodes' features the role, the number of films in which the cast component worked in the past, the average past income of the films in which the cast component worked in, the average critique vote of the films in which the cast component worked in, the information about past film genres and the current encoded role played by the cast component. Concluding, we remark how the assignment of the label to each team was obtained by discretizing through equally spaced quantiles (0-33, 34-66,67-100) the gross worlwide income.

\subsubsection{Dribbble}

We collected 8.35GB worth of data by scraping \url{dribbble.com}. The raw database is partitioned into several files: \textit{users.tsv}, \textit{followers.tsv}, \textit{shots.tsv}, \textit{likes.tsv}, \textit{comments.tsv} and \textit{skills.txt}. To foster reproducibility and further cooperation, the authors release the dataset under request.  In order to obtain the final working dataset, raw data was extensively preprocessed in order to solve a plethora of logical and structural inconsistencies\footnote{For more details, we refer to the code in the Github repository.}. By processing the raw files, we finally arrive at the definition of a consolidated relational database \textit{dribbble.db}.

\vspace{3mm}
\noindent
In order to frame the problem of team performance prediction as a graph representation learning task, we restructure the available data according to several simplifying assumptions:

\begin{itemize}
    \item Team composition is not directly accessible in the scraped data, therefore, we defined as user's "hiring date" the timestamp of the first shot of a user for a specific team. Besides, we set the "firing date" at 12 months after the date the user posted the last shot for that team; 
    \item To obtain meaningful internal team connectivity structure, we retained only users belonging to teams with at least three members;
    \item We considered teams that published at least one shot between 2017-01-01 and 2017-04-01. If a team published more than one shot, we aggregated its activity into a unique metric (i.e.: the average of the number of received likes in the defined time window).
    \item Firstly, the composition of the teams is built by looking for which members belonged to the team at the time of the publication of the shots. Secondly, it is kept an unique composition of the team considering the union of the members\footnote{Since a team may have published more than one shot then the composition of the team may have been different at the time of publication.}.
\end{itemize}

\vspace{3mm}

\noindent
The connectivity of nodes in the graph is defined by encoding "follow" users' interactions as edges. In more depth, we consider all of the "follow" actions up to 2017-04-01. The resulting directed graph is populated by $5\,196$ nodes and $304\,315$ edges. The nodes are partitioned into $769$ variable size and possibly overlapping teams. In order to define teams' labels we proceed as follows:

\begin{enumerate}
    \item We compute the average number of likes received by teams' contents in a 7-day from posting time interval;
    \item The obtained numbers are discretized into three classes according to equally spaced quantiles (0-33, 34-66,67-100).
\end{enumerate}

Concluding, we have considered as nodes' attributes some social features computed starting from 12-24-2016: number of received likes, number of shots published, number of comments published and received, number of skills and size of the team.

\subsubsection{Kaggle}

We downloaded the raw dataset from Kaggle website: \url{https://www.kaggle.com/kaggle/meta-kaggle}. In particular, we only leveraged information contained in several tables: \textit{Users.csv}, \textit{Teams.csv}, \textit{Competitions.csv}, \textit{TeamMemberships.csv}, \textit{ForumMessages.csv}, \textit{Kernels.csv} and \textit{UserFollowers.csv}. 

\noindent
In order to frame the problem of team performance prediction as a graph representation learning task, we restructure the available data according to several simplifying assumptions:

\begin{itemize}
    \item To obtain meaningful internal team connectivity structure, we retained only users belonging to teams with at least four members;
    \item We normalize the ranking position based on the size of the corresponding competition; 
    \item We take the average ranking position if a team partecipate to multiple competitions;
    \item We only considered the teams that participated in some competitions between 2020-11-01 and 2021-06-01 and ranked within the private leaderboard.
\end{itemize}

\noindent
The connectivity of nodes in the graph is defined by encoding "follow" users' interactions as edges. In more depth, we consider all of the "follow" actions up to 2021-06-11. The resulting directed graph is populated by $4\,183$ nodes and $17\,789$ edges. The nodes are partitioned into $1\,013$ variable size and possibly overlapping teams. We add an extra level of connectivity interconnecting the users that co-worked, consequently making the internal structure of team a clique. We considered as nodes' features the number of medals (gold, silver and bronze) that the users earned in 3 main categories (\textit{Competition}, \textit{Discussion} and \textit{Kernel}), the number of teamleader roles and the number of competitions before the reference starting date. Concluding, we remark how the assignment of the label to each team was obtained by
discretizing through equally spaced quantiles (0-33, 34-66,67-100) the average ranking position.

\section{Details on Empirical Evaluation}
\label{app:empirical_eval}

As described in Section \ref{sec:compared}, the classical machine learning methods work on hand-engineered features. In particular, we consider the following network features at subgraph level: total connections inside the team, number of unique and total followers outside the team, number of unique and total followings outside the team,  assortativity, density, average clustering and size of the team. Furthermore, the attributes of the nodes are added to the tabular data by means of an aggregation function, i.e., Aggr$_{team}$, in order to make possible the classification task at team level.

\vspace{3mm}
\noindent
In Tables \ref{tab:F1_AUROC_synt} and \ref{tab:F1_AUROC_real}, we show performance results using AUROC metric.

\begin{table*}[ht]
    \centering
    \begin{tabular}{m{1em} c c c c c c c c}
        &\\
        \hline
        & \makecell{Dataset} & \makecell{CIn} & \makecell{COut} & \makecell{T1} & \makecell{T2}& \makecell{T3} & \makecell{P}& \makecell{PT} \\
        \hline
        \rotatebox{90}{\textbf{Classical ML}}& \makecell{LR \\ SVM\\ RF \\ XGBoost \\ MLP} & \makecell{$100.0 \pm 0.0$ \\ $100.0 \pm 0.0$ \\ $100.0 \pm 0.1$ \\ $99.8 \pm 0.6$ \\ $100.0\pm 0.0 $} & \makecell{$100.0 \pm 0.0$ \\ $100.0 \pm 0.0$ \\ $100.0 \pm 0.0$ \\ $100.0 \pm 0.0$ \\ $100.0\pm0.0$} & \makecell{$61.4 \pm 2.9$ \\ $63.1 \pm 2.8$ \\ $62.8 \pm 2.6$ \\ $63.0 \pm 2.7$ \\ $63.7\pm2.0$} & \makecell{$65.3 \pm 3.8$ \\ $64.8 \pm 3.4$ \\ $64.8 \pm 3.4$ \\ $64.8 \pm 3.8$ \\ $65.1\pm3.4$}& \makecell{$66.5 \pm 2.8$ \\ $68.5 \pm 2.8$ \\ $66.0 \pm 2.7$ \\ $66.1 \pm 2.9$ \\ $69.1\pm2.9$} & \makecell{$50.0 \pm 0.0$ \\ $50.0 \pm 0.0$ \\ $50.0 \pm 0.0$ \\ $50.0 \pm 0.0$ \\ $50.0\pm0.0$}& \makecell{$71.9.4 \pm 1.5$ \\ $71.1 \pm 1.4$ \\ $71.8 \pm 2.3$ \\ $71.9 \pm 2.0$ \\ $70.9\pm1.9$} \\
        \hline
        \rotatebox{90}{\textbf{GNN}}& \makecell{SubGNN \\ MENTOR - T \\ MENTOR - C \\ MENTOR - L \\ MENTOR} & \makecell{$ 75.9\pm16.6 $ \\ $49.9 \pm 3.8$ \\ $100.0 \pm 0.0$ \\ $74.2 \pm 1.5$ \\ $100.0 \pm 0.0$} & \makecell{$77.2\pm16.5 $ \\ $51.1 \pm 2.7$ \\ $100.0 \pm 0.0$ \\ $91.8 \pm 6.4$ \\ $100.0 \pm 0.0$} & \makecell{$ 56.9\pm5.5 $ \\ $99.1 \pm 0.4$ \\ $49.5 \pm 3.4$ \\ $52.4 \pm 3.3$ \\ $99.1 \pm 0.4$} & \makecell{$53.6 \pm 9.3 $ \\ $99.3 \pm 0.4$ \\ $50.1 \pm 3.7$ \\ $52.1 \pm 2.9$ \\ $99.1 \pm 0.5$}& \makecell{$ 65.0\pm7.0 $ \\ $99.3 \pm 0.3$ \\ $59.4 \pm 6.3$ \\ $48.9 \pm 3.1$ \\ $99.1 \pm 0.4$} & \makecell{$ 99.5\pm1.0 $ \\ $49.3 \pm 1.6$ \\ $50.0 \pm 0.0$ \\ $100.0 \pm 0.0$ \\ $100.0 \pm 0.0$}& \makecell{$ 78.9\pm6.2 $ \\ $71.0 \pm 1.5$ \\ $60.9 \pm 1.4$ \\ $73.8 \pm 2.3$ \\ $96.6 \pm 0.9$} \\
        \hline
    \end{tabular}
    \label{tab:AUROC_1}
    \caption{\textbf{AUROC on synthetic datasets}. Standard deviations are provided from runs with 10 random seeds.}
    \label{tab:F1_AUROC_synt}
\end{table*}

\begin{table}[ht]
    \centering
    \begin{tabular}{m{1em} c c c c c}
        &\\
        \hline
        & \makecell{Dataset} & \makecell{IMDb} & \makecell{Dribbble} & \makecell{Kaggle}\\
        \hline
        \rotatebox{90}{\textbf{Classical ML}}& \makecell{LR \\ SVM\\ RF \\ XGBoost \\ MLP} & \makecell{$82.1 \pm 2.5$ \\ $82.1 \pm 2.5$ \\ $81.5 \pm 2.5$ \\ $81.5 \pm 2.2$ \\ $82.1 \pm 2.6$} & \makecell{$82.0 \pm 2.2$ \\ $83.2 \pm 2.1$ \\ $81.6 \pm 2.6$ \\ $81.4 \pm 2.3$ \\ $82.6\pm2.1$} & \makecell{$63.9 \pm 2.3$ \\ $63.0 \pm 2.3$ \\ $62.8 \pm 2.2$ \\ $62.9 \pm 2.3$ \\ $ 64.2\pm 2.6$}\\
        \hline
        \rotatebox{90}{\textbf{GNN}}& \makecell{SubGNN \\ MENTOR - T \\ MENTOR - C \\ MENTOR - L \\ MENTOR} & \makecell{$81.1 \pm 1.9$ \\ $83.7 \pm 4.2$ \\ $68.2 \pm 2.7$ \\ $69.6 \pm 2.0$ \\ $84.2 \pm 2.6$} & \makecell{$ 78.9\pm1.4 $ \\ $81.2 \pm 2.1$ \\ $81.5 \pm 2.7$ \\ $63.5 \pm 3.7$ \\ $83.3 \pm 2.7$} & \makecell{$ - $ \\ $64.1 \pm 2.0$ \\ $63.2 \pm 2.7$ \\ $63.1 \pm 3.1$ \\ $63.9 \pm 2.4$}\\
        \hline
    \end{tabular}
    \label{tab:AUROC_2}
    \caption{\textbf{AUROC on real-world datasets}. Standard deviations are provided from runs with 10 random seeds.}
    \label{tab:F1_AUROC_real}
\end{table}

\section{Implementation Details}
\label{app:implementation}

\textbf{Computing infrastructure.} We leverage Pytorch Geometric (Version 2.0.1) \cite{torch_geometric} and Pytorch Lightning (Version 1.3.1) \cite{falcon2019pytorch} for model development. Models were trained on single GPUs from a cluster containing NVIDIA 2070 RTX, NVIDIA GeForce GTX TITAN X and NVIDIA Quadro RTX 6000 GPUs.

\vspace{3mm}
\noindent
\textbf{Model hyperparameter tuning.} Hyperparameters were selected to optimize accuracy scores on the validation datasets (5-fold validation) using Optuna \cite{optuna}. In Table \ref{tab:hyper}, we describe the hyperparameter ranges we explored for each model. 

\begin{table}[ht]
      \centering
        \begin{tabular}{ll}
        \hline
        Parameter & Distribution \\ \hline
        Aggr$_{team}$      & Choice[sum, mean, max, min]         \\ 
        $C$      & Choice[0.1, 1, 10, 100, 1000]          \\ 
        Normalization      & Choice[MinMax, Standard, Robust, Quantile]          \\ \hline
        \# Iterations      & 200, Early Stopping 80          \\ \hline
        \end{tabular}
        \caption*{Logistic Regression}
\end{table}

\begin{table}[ht]
      \centering
        \begin{tabular}{ll}
        \hline
        Parameter & Distribution \\ \hline
        Aggr$_{team}$      & Choice[sum, mean, max, min]         \\ 
        $C$      & Choice[0.1, 1, 10, 100, 1000]          \\
        gamma      & Choice[1, 0.1, 0.01, 0.001, 0.0001]          \\
        Normalization      & Choice[MinMax, Standard, Robust, Quantile]          \\ \hline
        \# Iterations      & 200, Early Stopping 80          \\ \hline
        \end{tabular}
        \caption*{Support Vector Machine}
\end{table}

\begin{table}[ht]
      \centering
        \begin{tabular}{ll}
        \hline
        Parameter & Distribution \\ \hline
        Aggr$_{team}$      & Choice[sum, mean, max, min]         \\ 
        Criterion      & Choice[entropy, gini]          \\
        Max depth      & IntUniform[1, 40]          \\
        Min samples split      & IntUniform[1, 30]          \\
        Min samples leaf      & IntUniform[1, 40]          \\
        Max features      & Uniform[0.6, 1]          \\
        Max samples      & Uniform[0.6, 1]          \\
        Normalization      & Choice[MinMax, Standard, Robust, Quantile]          \\ \hline
        \# Iterations      & 200, Early Stopping 80          \\ \hline
        \end{tabular}
        \caption*{Random Forest}
 \end{table}
 
\begin{table}[ht]
 
      \centering
      \begin{tabular}{ll}
    \hline
    Parameter & Distribution \\ \hline
    Aggr$_{team}$      & Choice[sum, mean, max, min]         \\ 
    Criterion      & Choice[entropy, gini]          \\
    Max depth      & IntUniform[3, 10]          \\
    Min samples weight      & Uniform[1, 10]          \\
    Gamma      & Uniform[0, 1]          \\
    Reg. alpha      & Uniform[0, 1]          \\
    Reg. lambda      & Uniform[0, 1]          \\
    Learning rate      & Uniform[0.001, 0.3]          \\
    Subsample      & Uniform[0.6, 1]          \\
    Colsample bytree      & Uniform[0.6, 1]          \\
    Normalization      & Choice[MinMax, Standard, Robust, Quantile]          \\ \hline
    \# Iterations      & 200, Early Stopping 80          \\ \hline
    \end{tabular}
    \caption*{XGBoost}
\end{table}

\begin{table}[ht]

      \centering
        \begin{tabular}{ll}
    \hline
    Parameter & Distribution \\ \hline
    Aggr$_{team}$      & Choice[sum, mean, max, min]         \\ 
    Dropout      & DiscreteUniform[0.2, 0.8, 0.05]          \\ 
    Epochs      & DiscreteUniform[20, 100, 2]          \\ 
    Learning rate      & LogUniform[1e-5, 1e-1]          \\ 
    Num. layers      & IntUniform[1, 3]          \\ 
    Hidden dimension      & Choice[16, 32, 64, 128]          \\ 
    Normalization      & Choice[MinMax, Standard, Robust, Quantile]          \\ \hline
    \# Iterations      & 200, Early Stopping 80          \\ \hline
    \end{tabular}
    \caption*{Multi-layer Perceptron}
    \end{table}
    
\begin{table}[ht]
    
      \centering
      \begin{tabular}{ll}
    \hline
    Parameter & Distribution \\ \hline
    Aggr$_{conv}^{(T)}$      & Choice[sum, mean, max, min]        \\ 
    Aggr$_{conv}^{(C)}$      & Choice[sum, mean, max, min]          \\ 
    Aggr$_{conv}^{(P)}$      & Choice[sum, mean, max, min]          \\ 
    Aggr$_{team}^{(T)}$      & Choice[sum, mean, max, min]         \\ 
    Flow$_{conv}^{(T)}$      & Choice[source to target, target to source]          \\ 
    Flow$_{conv}^{(C)}$      & Choice[source to target, target to source]           \\ 
    Dropout$^{(T)}$      & DiscreteUniform[0.2, 0.8, 0.05]          \\ 
    Dropout$^{(C)}$      & DiscreteUniform[0.2, 0.8, 0.05]          \\ 
    Dropout$^{(A)}$      & DiscreteUniform[0.2, 0.8, 0.05]          \\ 
    Epochs      & DiscreteUniform[20, 100, 2]          \\ 
    Learning rate      & LogUniform[1e-5, 1e-1]          \\ 
    Learning rate (SWA)     & LogUniform[1e-5, 1e-1]          \\ 
    Start (SWA)      & DiscreteUniform[0.6, 0.95, 0.05]          \\ 
    Freq. (SWA)      & IntUniform[1, 20]          \\ 
    Hidden dimension      & Choice[16, 32, 64, 128]          \\ 
    Normalization      & Choice[MinMax, Standard, Robust, Quantile]          \\ \hline
    \# Iterations      & 200, Early Stopping 80          \\ \hline
    \end{tabular}
    \caption*{(f)}
    \caption{\textbf{Hyperparameter ranges} MENTOR (3-channels and 1-channel)}
    \label{tab:hyper}
\end{table}

\end{appendix}


\bibliographystyle{abbrv}  
\bibliography{ms}

\begin{thebibliography}{10}

\bibitem{gini1}
R.~Abraham, S.~Bergh, and P.~Nair.
\newblock A new approach to galaxy morphology: I. analysis of the sloan digital
  sky survey early data release.
\newblock {\em The Astrophysical Journal}, 588, 01 2003.

\bibitem{optuna}
T.~Akiba, S.~Sano, T.~Yanase, T.~Ohta, and M.~Koyama.
\newblock Optuna: A next-generation hyperparameter optimization framework.
\newblock In {\em Proceedings of the 25th ACM SIGKDD international conference
  on knowledge discovery \& data mining}, pages 2623--2631, 2019.

\bibitem{SUBGNN}
E.~Alsentzer, S.~G. Finlayson, M.~M. Li, and M.~Zitnik.
\newblock Subgraph neural networks.
\newblock {\em arXiv preprint arXiv:2006.10538}, 2020.

\bibitem{impact}
B.~M. Althouse, J.~D. West, C.~T. Bergstrom, and T.~Bergstrom.
\newblock Differences in impact factor across fields and over time.
\newblock {\em Journal of the American Society for Information Science and
  Technology}, 60(1):27--34, 2009.

\bibitem{arrow}
H.~Arrow, J.~McGrath, and J.~Berdahl.
\newblock {\em Small Groups As Complex Systems: Formation, Coordination,
  Development, and Adaptation}.
\newblock 01 2000.

\bibitem{barabasi1999emergence}
A.-L. Barab{\'a}si and R.~Albert.
\newblock Emergence of scaling in random networks.
\newblock {\em science}, 286(5439):509--512, 1999.

\bibitem{battaglia2018relational}
P.~W. Battaglia, J.~B. Hamrick, V.~Bapst, A.~Sanchez-Gonzalez, V.~Zambaldi,
  M.~Malinowski, A.~Tacchetti, D.~Raposo, A.~Santoro, R.~Faulkner, et~al.
\newblock Relational inductive biases, deep learning, and graph networks.
\newblock {\em arXiv preprint arXiv:1806.01261}, 2018.

\bibitem{Bell2007}
S.~Bell.
\newblock Deep-level composition variables as predictors of team performance.
\newblock {\em The Journal of applied psychology}, 92:595--615, 06 2007.

\bibitem{bell2018}
S.~Bell, S.~Brown, A.~Colaneri, and N.~Outland.
\newblock Team composition and the abcs of teamwork.
\newblock {\em American Psychologist}, 73:349--362, 05 2018.

\bibitem{Bell2011}
S.~Bell, A.~Villado, M.~Lukasik, L.~Belau, and A.~Briggs.
\newblock Getting specific about demographic diversity variable and team
  performance relationships: A meta-analysis.
\newblock {\em Journal of Management}, 37:709--743, 05 2011.

\bibitem{tpe}
J.~Bergstra, R.~Bardenet, Y.~Bengio, and B.~K{\'e}gl.
\newblock Algorithms for hyper-parameter optimization.
\newblock {\em Advances in neural information processing systems}, 24, 2011.

\bibitem{tpe2}
J.~Bergstra, D.~Yamins, and D.~Cox.
\newblock Making a science of model search: Hyperparameter optimization in
  hundreds of dimensions for vision architectures.
\newblock In {\em International conference on machine learning}, pages
  115--123. PMLR, 2013.

\bibitem{gini3}
L.~Bertoli-Barsotti and T.~Lando.
\newblock How mean rank and mean size may determine the generalised lorenz
  curve: With application to citation analysis.
\newblock {\em Journal of Informetrics}, 13(1):387--396, 2019.

\bibitem{bo2021beyond}
D.~Bo, X.~Wang, C.~Shi, and H.~Shen.
\newblock Beyond low-frequency information in graph convolutional networks.
\newblock {\em arXiv preprint arXiv:2101.00797}, 2021.

\bibitem{borgatti}
S.~Borgatti and P.~Foster.
\newblock The network paradigm in organizational research: A review and
  typology.
\newblock {\em Journal of Management}, 29:991--1013, 12 2003.

\bibitem{Brannick1997TeamPA}
M.~T. Brannick, E.~Salas, and C.~Prince.
\newblock Team performance assessment and measurement: Theory, methods, and
  applications. series in applied psychology.
\newblock 1997.

\bibitem{breiman2001random}
L.~Breiman.
\newblock Random forests.
\newblock {\em Machine learning}, 45(1):5--32, 2001.

\bibitem{GATv2}
S.~Brody, U.~Alon, and E.~Yahav.
\newblock How attentive are graph attention networks?
\newblock {\em CoRR}, abs/2105.14491, 2021.

\bibitem{carter2015little}
D.~R. Carter, R.~Asencio, A.~Wax, L.~A. DeChurch, and N.~S. Contractor.
\newblock Little teams, big data: Big data provides new opportunities for teams
  theory.
\newblock {\em Industrial and Organizational Psychology}, 8(4):550--555, 2015.

\bibitem{chami2020machine}
I.~Chami, S.~Abu-El-Haija, B.~Perozzi, C.~R{\'e}, and K.~Murphy.
\newblock Machine learning on graphs: A model and comprehensive taxonomy.
\newblock {\em arXiv preprint arXiv:2005.03675}, 2020.

\bibitem{Chen2004}
G.~Chen, J.~Mathieu, and P.~Bliese.
\newblock A framework for conducting multilevel construct validation.
\newblock {\em Research in Multi Level Issues}, 3:273--303, 12 2003.

\bibitem{chen2016xgboost}
T.~Chen and C.~Guestrin.
\newblock Xgboost: A scalable tree boosting system.
\newblock In {\em Proceedings of the 22nd acm sigkdd international conference
  on knowledge discovery and data mining}, pages 785--794, 2016.

\bibitem{Chen2019}
Z.~Cheng, Y.~Yang, C.~Tan, D.~Cheng, A.~Cheng, and Y.~Zhuang.
\newblock What makes a good team? a large-scale study on the effect of team
  composition in honor of kings.
\newblock pages 2666--2672, 05 2019.

\bibitem{gini2}
A.~Delbosc and G.~Currie.
\newblock Using lorenz curves to assess public transport equity.
\newblock {\em Journal of Transport Geography}, 19(6):1252--1259, 2011.
\newblock Special section on Alternative Travel futures.

\bibitem{delice2019advancing}
F.~Delice, M.~Rousseau, and J.~Feitosa.
\newblock Advancing teams research: What, when, and how to measure team
  dynamics over time.
\newblock {\em Frontiers in psychology}, 10:1324, 2019.

\bibitem{ducanis1979interdisciplinary}
A.~Ducanis and A.~Golin.
\newblock {\em The Interdisciplinary Health Care Team: A Handbook}.
\newblock Aspen publication. Aspen Systems Corporation, 1979.

\bibitem{NYT}
C.~Duhigg.
\newblock What google learned from its quest to build the perfect team.
\newblock {\em The New York Times Magazine}, 2016.

\bibitem{erdos1960evolution}
P.~Erdos, A.~R{\'e}nyi, et~al.
\newblock On the evolution of random graphs.
\newblock {\em Publ. Math. Inst. Hung. Acad. Sci}, 5(1):17--60, 1960.

\bibitem{falcon2019pytorch}
W.~{Falcon et al.}
\newblock Pytorch lightning.
\newblock {\em GitHub. Note:
  https://github.com/PyTorchLightning/pytorch-lightning}, 3, 2019.

\bibitem{torch_geometric}
M.~Fey and J.~E. Lenssen.
\newblock Fast graph representation learning with pytorch geometric.
\newblock {\em CoRR}, abs/1903.02428, 2019.

\bibitem{forsyth2009group}
D.~Forsyth.
\newblock {\em Group Dynamics}.
\newblock Cengage Learning, 2009.

\bibitem{gao2019edge2vec}
Z.~Gao, G.~Fu, C.~Ouyang, S.~Tsutsui, X.~Liu, J.~Yang, C.~Gessner, B.~Foote,
  D.~Wild, Y.~Ding, et~al.
\newblock edge2vec: Representation learning using edge semantics for biomedical
  knowledge discovery.
\newblock {\em BMC bioinformatics}, 20(1):1--15, 2019.

\bibitem{gilmer2017neural}
J.~Gilmer, S.~S. Schoenholz, P.~F. Riley, O.~Vinyals, and G.~E. Dahl.
\newblock Neural message passing for quantum chemistry.
\newblock In {\em International conference on machine learning}, pages
  1263--1272. PMLR, 2017.

\bibitem{godwin2021very}
J.~Godwin, M.~Schaarschmidt, A.~Gaunt, A.~Sanchez-Gonzalez, Y.~Rubanova,
  P.~Veli{\v{c}}kovi{\'c}, J.~Kirkpatrick, and P.~Battaglia.
\newblock Very deep graph neural networks via noise regularisation.
\newblock {\em arXiv preprint arXiv:2106.07971}, 2021.

\bibitem{goyal}
P.~Goyal, A.~Sapienza, and E.~Ferrara.
\newblock Recommending teammates with deep neural networks.
\newblock pages 57--61, 07 2018.

\bibitem{grover2016node2vec}
A.~Grover and J.~Leskovec.
\newblock node2vec: Scalable feature learning for networks.
\newblock In {\em Proceedings of the 22nd ACM SIGKDD international conference
  on Knowledge discovery and data mining}, pages 855--864, 2016.

\bibitem{guimera}
R.~Guimerà, B.~Uzzi, J.~Spiro, and L.~Amaral.
\newblock Team assembly mechanisms determine collaboration network structure
  and team performance.
\newblock {\em Science (New York, N.Y.)}, 308:697--702, 05 2005.

\bibitem{GraphSAGE}
W.~Hamilton, R.~Ying, and J.~Leskovec.
\newblock Inductive representation learning on large graphs.
\newblock 06 2017.

\bibitem{hamilton2020graph}
W.~L. Hamilton.
\newblock Graph representation learning.
\newblock {\em Synthesis Lectures on Artifical Intelligence and Machine
  Learning}, 14(3):1--159, 2020.

\bibitem{resnet}
K.~He, X.~Zhang, S.~Ren, and J.~Sun.
\newblock Deep residual learning for image recognition.
\newblock In {\em Proceedings of the IEEE conference on computer vision and
  pattern recognition}, pages 770--778, 2016.

\bibitem{Humphrey}
S.~Humphrey, J.~HOLLENBECK, C.~Meyer, and D.~Ilgen.
\newblock Personality configurations in self‐managed teams: A natural
  experiment on the effects of maximizing and minimizing variance in traits.
\newblock {\em Journal of Applied Social Psychology}, 41:1701 -- 1732, 07 2011.

\bibitem{swa}
P.~Izmailov, D.~Podoprikhin, T.~Garipov, D.~Vetrov, and A.~G. Wilson.
\newblock Averaging weights leads to wider optima and better generalization.
\newblock {\em arXiv preprint arXiv:1803.05407}, 2018.

\bibitem{adam}
D.~P. Kingma and J.~Ba.
\newblock Adam: A method for stochastic optimization.
\newblock {\em arXiv preprint arXiv:1412.6980}, 2014.

\bibitem{GCN}
T.~N. Kipf and M.~Welling.
\newblock Semi-supervised classification with graph convolutional networks.
\newblock {\em arXiv preprint arXiv:1609.02907}, 2016.

\bibitem{klicpera2018predict}
J.~Klicpera, A.~Bojchevski, and S.~G{\"u}nnemann.
\newblock Predict then propagate: Graph neural networks meet personalized
  pagerank.
\newblock {\em arXiv preprint arXiv:1810.05997}, 2018.

\bibitem{kossinets2009origins}
G.~Kossinets and D.~J. Watts.
\newblock Origins of homophily in an evolving social network.
\newblock {\em American journal of sociology}, 115(2):405--450, 2009.

\bibitem{Kozlowski}
S.~Kozlowski and K.~Klein.
\newblock A multilevel approach to theory and research in organizations:
  Contextual, temporal, and emergent processes.
\newblock {\em Multi-level theory, research, and methods in organizations:
  Foundations, extensions, and new directions}, 10 2012.

\bibitem{levine2006small}
J.~Levine and R.~Moreland.
\newblock {\em Small Groups: Key Readings}.
\newblock Key readings in social psychology. Psychology Press, 2006.

\bibitem{edges2}
Q.~Li, Z.~Cao, J.~Zhong, and Q.~Li.
\newblock Graph representation learning with encoding edges.
\newblock {\em Neurocomputing}, 361, 07 2019.

\bibitem{li2018deeper}
Q.~Li, Z.~Han, and X.-M. Wu.
\newblock Deeper insights into graph convolutional networks for semi-supervised
  learning.
\newblock In {\em Thirty-Second AAAI conference on artificial intelligence},
  2018.

\bibitem{globalattention}
Y.~Li, D.~Tarlow, M.~Brockschmidt, and R.~Zemel.
\newblock Gated graph sequence neural networks.
\newblock {\em arXiv preprint arXiv:1511.05493}, 2015.

\bibitem{liao2018attributed}
L.~Liao, X.~He, H.~Zhang, and T.-S. Chua.
\newblock Attributed social network embedding.
\newblock {\em IEEE Transactions on Knowledge and Data Engineering},
  30(12):2257--2270, 2018.

\bibitem{maron2019provably}
H.~Maron, H.~Ben-Hamu, H.~Serviansky, and Y.~Lipman.
\newblock Provably powerful graph networks.
\newblock {\em arXiv preprint arXiv:1905.11136}, 2019.

\bibitem{Mathieu}
J.~Mathieu, S.~Tannenbaum, J.~Donsbach, and Alliger.
\newblock A review and integration of team composition models: Moving toward a
  dynamic and temporal framework.
\newblock {\em Journal of Management}, 40:130--160, 12 2013.

\bibitem{McGrath}
J.~E. McGrath.
\newblock {\em Social psychology : a brief introduction / Joseph E. McGrath}.
\newblock Holt, Rinehart and Winston New York, 1964.

\bibitem{mcpherson2001birds}
M.~McPherson, L.~Smith-Lovin, and J.~M. Cook.
\newblock Birds of a feather: Homophily in social networks.
\newblock {\em Annual review of sociology}, 27(1):415--444, 2001.

\bibitem{merton1968matthew}
R.~K. Merton.
\newblock The matthew effect in science: The reward and communication systems
  of science are considered.
\newblock {\em Science}, 159(3810):56--63, 1968.

\bibitem{micheli2009neural}
A.~Micheli.
\newblock Neural network for graphs: A contextual constructive approach.
\newblock {\em IEEE Transactions on Neural Networks}, 20(3):498--511, 2009.

\bibitem{Peeters}
M.~Peeters, H.~Tuijl, C.~Rutte, and I.~Reymen.
\newblock Personality and team performance: A meta-analysis.
\newblock {\em European Journal of Personality}, 20:377 -- 396, 08 2006.

\bibitem{pentland2012new}
A.~S. Pentland.
\newblock The new science of building great teams.
\newblock {\em Harvard business review}, 90(4):60--69, 2012.

\bibitem{perozzi2014deepwalk}
B.~Perozzi, R.~Al-Rfou, and S.~Skiena.
\newblock Deepwalk: Online learning of social representations.
\newblock In {\em Proceedings of the 20th ACM SIGKDD international conference
  on Knowledge discovery and data mining}, pages 701--710, 2014.

\bibitem{ramos}
P.~J. Ramos-Villagrasa, P.~Marques-Quinteiro, J.~Navarro, and R.~Rico.
\newblock Teams as complex adaptive systems: Reviewing 17 years of research.
\newblock {\em Small Group Research}, 49, 05 2017.

\bibitem{sabidussi1966centrality}
G.~Sabidussi.
\newblock The centrality index of a graph.
\newblock {\em Psychometrika}, 31(4):581--603, 1966.

\bibitem{sapienza}
A.~Sapienza, P.~Goyal, and E.~Ferrara.
\newblock Deep neural networks for optimal team composition.
\newblock {\em CoRR}, abs/1805.03285, 2018.

\bibitem{scarselli2008graph}
F.~Scarselli, M.~Gori, A.~C. Tsoi, M.~Hagenbuchner, and G.~Monfardini.
\newblock The graph neural network model.
\newblock {\em IEEE transactions on neural networks}, 20(1):61--80, 2008.

\bibitem{srinivasan2019equivalence}
B.~Srinivasan and B.~Ribeiro.
\newblock On the equivalence between positional node embeddings and structural
  graph representations.
\newblock {\em arXiv preprint arXiv:1910.00452}, 2019.

\bibitem{uzzi}
B.~Uzzi and J.~Spiro.
\newblock Collaboration and creativity: The small world problem.
\newblock {\em American Journal of Sociology - AMER J SOCIOL}, 111:447--504, 09
  2005.

\bibitem{allyouneed}
A.~Vaswani, N.~Shazeer, N.~Parmar, J.~Uszkoreit, L.~Jones, A.~N. Gomez,
  L.~Kaiser, and I.~Polosukhin.
\newblock Attention is all you need.
\newblock {\em arXiv preprint arXiv:1706.03762}, 2017.

\bibitem{GAT}
P.~Veličković, G.~Cucurull, A.~Casanova, A.~Romero, P.~Lio, and Y.~Bengio.
\newblock Graph attention networks.
\newblock 10 2017.

\bibitem{caveman}
D.~J. Watts.
\newblock Networks, dynamics, and the small-world phenomenon.
\newblock {\em American Journal of sociology}, 105(2):493--527, 1999.

\bibitem{SurveyGNN}
Z.~Wu, S.~Pan, F.~Chen, G.~Long, C.~Zhang, and P.~S. Yu.
\newblock A comprehensive survey on graph neural networks.
\newblock {\em CoRR}, abs/1901.00596, 2019.

\bibitem{GIN}
K.~Xu, W.~Hu, J.~Leskovec, and S.~Jegelka.
\newblock How powerful are graph neural networks?
\newblock 10 2018.

\bibitem{jumping}
K.~Xu, C.~Li, Y.~Tian, T.~Sonobe, K.-i. Kawarabayashi, and S.~Jegelka.
\newblock Representation learning on graphs with jumping knowledge networks.
\newblock In {\em International Conference on Machine Learning}, pages
  5453--5462. PMLR, 2018.

\bibitem{xu2019can}
K.~Xu, J.~Li, M.~Zhang, S.~S. Du, K.-i. Kawarabayashi, and S.~Jegelka.
\newblock What can neural networks reason about?
\newblock {\em arXiv preprint arXiv:1905.13211}, 2019.

\bibitem{yan2021two}
Y.~Yan, M.~Hashemi, K.~Swersky, Y.~Yang, and D.~Koutra.
\newblock Two sides of the same coin: Heterophily and oversmoothing in graph
  convolutional neural networks.
\newblock {\em arXiv preprint arXiv:2102.06462}, 2021.

\bibitem{you2019position}
J.~You, R.~Ying, and J.~Leskovec.
\newblock Position-aware graph neural networks.
\newblock In {\em International Conference on Machine Learning}, pages
  7134--7143. PMLR, 2019.

\bibitem{PGNN}
J.~You, R.~Ying, and J.~Leskovec.
\newblock Position-aware graph neural networks.
\newblock 06 2019.

\bibitem{yucesoy2016untangling}
B.~Yucesoy and A.-L. Barab{\'a}si.
\newblock Untangling performance from success.
\newblock {\em EPJ Data Science}, 5(1):1--10, 2016.

\end{thebibliography}

\end{document}